\RequirePackage{ifpdf}
\documentclass[preprint, preprintnumbrs]{article}
\usepackage{jheppub}
\usepackage{xcolor}
\usepackage{amsmath}
\usepackage{epsfig}
\usepackage{caption}
\usepackage{subcaption}
\usepackage{environ}
\usepackage{comment}
\usepackage{soul}
\pdfoutput=1

\newcommand{\roughly}[1]{\mathrel{\raise.3ex\hbox{$#1$\kern-0.85em
\lower1ex\hbox{$\sim$}}}}

\def\mc{\mathcal}
\def\nn{\nonumber}

\newcommand{\be}{\begin{equation}}
\newcommand{\bee}{\begin{equation}}
\newcommand{\ee}{\end{equation}}
\newcommand{\beea}{\begin{eqnarray}}
\newcommand{\eea}{\end{eqnarray}}
\newcommand{\bea}{\begin{eqnarray}}
\newcommand{\ea}{\end{eqnarray}}
\newcommand{\ba}{\begin{eqnarray}}
\newcommand{\eq}{\begin{equation}}
\newcommand{\eeq}{\end{equation}}
\newcommand{\eqa}{\begin{eqnarray}}
\newcommand{\eeqa}{\end{eqnarray}}

 \NewEnviron{ignore}{}{}  

\def\nott#1{\setbox0=\hbox{$#1$}                
   \dimen0=\wd0                                 
   \setbox1=\hbox{/} \dimen1=\wd1               
   \ifdim\dimen0>\dimen1                        
      \rlap{\hbox to \dimen0{\hfil/\hfil}}      
      #1                                        
   \else                                        
      \rlap{\hbox to \dimen1{\hfil$#1$\hfil}}   
      /                                         
   \fi}                                         %

\def\uxsl{\hbox{/\kern-.4000em$u$}}
\def\uxslsm{\hbox{\smaller/\kern-.5600em$u$}}
\def\pxpsl{\hbox{/\kern-.5000em$p$}}
\def\epssl{\hbox{/\kern-.5600em$\epsilon$}}
\def\delsl{\hbox{/\kern-.7000em$\nabla$}}
\def\lxpsl{\hbox{/\kern-.5600em$l$}}
\def\kxpsl{\hbox{/\kern-.5600em$k$}}
\def\qxpsl{\hbox{/\kern-.3900em$q$}}

\def\eff{{\rm eff}}

\def\pref#1{(\ref{#1})}
\def\exd{{\rm d}}
\def\ol#1{{\overline{#1}}}

\def\Mp{M_p}

\def\cA{{\cal A}}
\def\cB{{\cal B}}
\def\cD{{\cal D}}
\def\cC{{\cal C}}
\def\cE{{\cal E}}
\def\cF{{\cal F}}
\def\cG{{\cal G}}
\def\cH{{\cal H}}

\def\cJ{{\cal J}}

\def\cL{{\cal L}}

\def\cN{{\cal N}}
\def\cO{{\cal O}}

\def\cR{{\cal R}}

\def\cT{{\cal T}}

\def\cZ{{\cal Z}}

\def\bfv{{\bf v}}

\def\bfx{{\bf x}}

\def\bfE{{\bf E}}
\def\bfJ{{\bf J}}

\def\bfU{{\bf U}}

\def\mfa{{\mathfrak a}}

\def\mfg{{\mathfrak g}}

\def\mfm{{\mathfrak m}}
\def\mfn{{\mathfrak n}}

\def\mfv{{\mathfrak v}}
\def\mfw{{\mathfrak w}}

\def\mfC{{\mathfrak C}}

\def\ssA{{\scriptscriptstyle A}}
\def\ssB{{\scriptscriptstyle B}}

\def\ssD{{\scriptscriptstyle D}}

\def\ssF{{\scriptscriptstyle F}}

\def\ssM{{\scriptscriptstyle M}}
\def\ssN{{\scriptscriptstyle N}}
\def\ssP{{\scriptscriptstyle P}}
\def\ssQ{{\scriptscriptstyle Q}}
\def\ssR{{\scriptscriptstyle R}}
\def\ssS{{\scriptscriptstyle S}}

\def\UV{{\scriptscriptstyle U\hbox{\kern-0.1em}V}}
\def\PPN{{\scriptscriptstyle P\hbox{\kern-0.1em}P\hbox{\kern-0.1em}N}}
\def\MN{{\scriptscriptstyle M\hbox{\kern-0.1em}N}}
\def\MNP{{\scriptscriptstyle M\hbox{\kern-0.1em}N\hbox{\kern-0.1em}P}}
\def\KK{{\scriptscriptstyle K\hbox{\kern-0.1em}K}}
\def\SM{{\scriptscriptstyle S\hbox{\kern-0.1em}M}}
\def\EH{{\scriptscriptstyle E\hbox{\kern-0.1em}H}}

\def\QCD{{\scriptscriptstyle Q\hbox{\kern-0.1em}C\hbox{\kern-0.1em}D}}
\def\IR{{\scriptscriptstyle I\hbox{\kern-0.1em}R}}
\def\TEV{{\scriptscriptstyle T\hbox{\kern-0.1em}E\hbox{\kern-0.1em}V}}

\def\Tr{{\rm Tr}}

\def\EW{{\scriptscriptstyle EW}}

\def\UV{{\scriptscriptstyle U\hbox{\kern-0.1em}V}}
\def\PPN{{\scriptscriptstyle P\hbox{\kern-0.1em}P\hbox{\kern-0.1em}N}}
\def\MN{{\scriptscriptstyle M\hbox{\kern-0.1em}N}}
\def\MNP{{\scriptscriptstyle M\hbox{\kern-0.1em}N\hbox{\kern-0.1em}P}}
\def\KK{{\scriptscriptstyle K\hbox{\kern-0.1em}K}}
\def\SM{{\scriptscriptstyle S\hbox{\kern-0.1em}M}}
\def\EH{{\scriptscriptstyle E\hbox{\kern-0.1em}H}}

\def\QCD{{\scriptscriptstyle Q\hbox{\kern-0.1em}C\hbox{\kern-0.1em}D}}
\def\IR{{\scriptscriptstyle I\hbox{\kern-0.1em}R}}
\def\TEV{{\scriptscriptstyle T\hbox{\kern-0.1em}E\hbox{\kern-0.1em}V}}
\def\aff{{a\hbox{\kern-0.1em}f\hbox{\kern-0.1em}f}}
\def\axion{{\mfa}}

\title{Dual is Different: EFTs, Axions and \\Nonpropagating Form Fields in Cosmology}

\author[a,b,c]{C.P.~Burgess}
\author[d, e]{and F. Quevedo}

\affiliation[a]{Department of Physics \& Astronomy, 
McMaster University, Hamilton, ON, Canada, L8S 4M1}

\affiliation[b]{Perimeter Institute for Theoretical Physics, 
Waterloo, ON, Canada, N2L 2Y5}

\affiliation[c]{School of Theoretical Physics, Dublin Institute for
 Advanced Studies, 10 Burlington Rd., 
 Dublin, Ireland}

\affiliation[d]{DAMTP, University of Cambridge, Wilberforce Road,  Cambridge, CB3 0WA, UK}

\affiliation[e]{New York University Abu Dhabi, 
Saadiyat Island, Abu Dhabi, United Arab Emirates}

\emailAdd{cburgess@perimeterinstitute.ca}
\emailAdd{fq201@cam.ac.uk}

\date{today}

\abstract{Scalar fields in 4D are known to have equivalent dual descriptions in terms of form-field gauge potentials, but this is often regarded as an arcane fact. Why use more complicated formulations when simpler scalar descriptions exist and are equivalent? We describe three ways in which scalars that arise as duals can differ from their garden-variety counterparts. Two of these -- the interchange of weak and strong couplings and utility in bringing topological information to the low-energy theory -- are relatively well-known, but to these we add a third: dualities that map derivative interactions to non-derivative interactions seem not to commute with the general power-counting arguments that quantify control over the low-energy approximation within any EFT involving gravity. Since both sides of the duality must agree on their low-energy limit, the non-derivative interactions on the scalar side of the duality turn out to be suppressed relative to what would generically be assumed. They are nonetheless technically natural, as is particularly clear in the dual formulation. We argue that scalar fields arising as duals (such as the universal axion $\mfa$ from string vacua) that have the commonly assumed $J^\mu \partial_\mu \axion$ interaction with matter also have $J^\mu J_\mu$ contact interactions among the respective currents, some of whose implications we explore. We also emphasize the non-trivial role and cosmological implications of non-propagating 3-form gauge potentials and clarify confusing statements recently made in the  literature regarding the validity of duality for massive form fields.
} 

\begin{document}
\maketitle

\section{Introduction}

Among the bigger theoretical realizations of the past decades is the relatively recent appreciation of how much smaller the space of quantum field theories (QFTs) is compared to what had been expected: what had been thought to have been distinct theories have often been found instead to be equivalent to one another. This equivalence is usually based on an explicit -- though sometimes counter-intuitive (in some cases the dual theories live in different spacetime dimensions) -- change of variables called a `duality' transformation (for a review see \cite{Polchinski:2014mva, Quevedo:1997jb}). Although physical predictions are identical for dual theories, their equivalence was sometimes hidden because the duality mapped weak to strong coupling and so was masked by the difficulty making explicit predictions in strongly coupled theories. 

Since this discovery the list of dual field theories has only grown. Among the simplest examples of such dualities are electric-magnetic types of dualities that in $D$ spacetime dimensions relate theories involving $p$-form gauge potentials to theories involving $q$-form gauge potentials with $q=D-p-2$. This boils down to electric-magnetic interchange in the special case of an electromagnetic potential in $D=4$ dimensions (for which $p=q=1$), but gives a variety of other dualities in other cases. Another $D = 4$ example is the duality between a 2-form Kalb Ramond potential $B_{\mu\nu}$ \cite{Kalb:1974yc} and a massless derivatively coupled scalar field (for which $p = 2$ and $q = 0$). 

Although often associated with massless fields these dualities  also extend to include massive particles \cite{Quevedo:1996uu, MassiveDual}, which on the form side acquire their masses through a generalization of the Stuckelberg mechanism. In this case it is a massive $p$ form dual to a massive $q=D-p-1$ form. In 4D there are at least two ways a 2-form potential $B_{\mu\nu}$ can acquire mass in this way.\footnote{The literature also contains mis-statements about what these massive dualities are -- {\it e.g.}~ref.~\cite{Hell:2021wzm} -- but we emphasize we mean the standard dualities whose validity is not in question when we here argue that dual is `different'.} In one of these the 2-form field is `eaten' by a Maxwell field $A_\mu$ in a dual version of the ordinary Higgs mechanism, leading to a massive spin-1 particle. In the other, the Kalb-Ramond field acquires a mass by `eating' a 3-form gauge potential, leading to a massive spin-0 particle. (This last version preserves the total number of degrees of freedom because for $p = 3$ in $D=4$ dimension we have $q = -1$; that is a 3-form field turns out not to propagate any physical degree of freedom at all.)

A common attitude towards the 2-form/scalar duality in 4D is that it just provides a complicated way to think about what is otherwise a simple thing: a scalar field. It is nice to know they are equivalent, but because they are one can be happy to stick with the simple formulation (thank you very much). A similar attitude applies to 3-forms -- why include them at all if they do not describe a propagating particle? 

Weak-strong coupling -- when duality maps a small coupling $g \ll 1$ on one side of the duality onto a large coupling $g \gg 1$ on the other -- provides one response to this attitude. If it is $g \ll 1$ (for some dimensionless coupling $g$) that provides control over the semiclassical expansion on the weakly coupled side then the  small-$g$ expansion breaks down in its dual. Although the classical solutions on both sides might be equivalent, all things being equal there is no reason to believe that this classical solution is actually a good approximation to what actually happens on the strong-coupling side.\footnote{There are many examples of this in string theory, such as in the duality between 11D supergravity (describing the low-energy limit of strongly coupled strings) and weakly coupled 10D F-theory vacua  (for a review see \cite{Denef:2008wq, Grimm:2013fua}). Although classical solutions on both sides of this duality map into one another, it is only on the weakly coupled 10D side that this classical solution is believed to be a good approximation to the full theory.} But duality does not always map strong to weak couplings, and when it doesn't it is common for practical people not to care much if a dual description exists.

The message of this paper is this: for low-energy effective field theories (EFTs) involving gravity there is a second way for duality to matter. It is often the low-energy expansion itself that controls the semiclassical approximation in gravity \cite{Weinberg:1978kz} (and for gravity coupled to other light fields \cite{Burgess:2009ea, Adshead:2017srh}), rather than expansions involving small dimensionless couplings (see \cite{Burgess:2003jk, Burgess:2020tbq} for reviews). Since the low-energy limit applies equally well on either side of the duality one might argue that in this case the classical approximation should be equally good on both sides. Although this is indeed true, in dualities involving massive scalar fields there can be a subtlety: the duality maps the non-derivative couplings of the scalar potential into couplings of the dual form fields that involve many derivatives. 

In this sense the duality mixes up different orders of the low-energy expansion, and as a result consistency of the low-energy approximation requires the interactions within the scalar potential on the scalar side -- which are the ones that are dangerous for the low-energy expansion -- to be systematically smaller than what would naively be called `generic'. For instance, the scalar potential generated by duality for a low-energy scalar $\axion$ with small mass $m$ turns out to be a function of the product $m \axion$ rather than $\axion$ alone: $V = V(m \axion)$ rather than $V(\axion)$ \cite{Burgess:2023ifd}. For instance a quartic interaction $\lambda \axion^4 \in V(\axion)$ actually arises under duality as $(m \axion)^4/M^4$ with some other scale $M$ providing the missing dimensions. In practice $M$ is usually the UV scale which on the form side suppressed interactions involving many derivatives, implying $\lambda \sim (m/M)^4 \ll 1$ whenever the scalar is light enough to be included within the low-energy theory. These small couplings are technically natural, and this is most easily seen by powercounting on the dual side where they correspond to ordinary higher-derivative interactions.\footnote{There can also be cases where $M$ is not large, such as for the dual of the QCD axion where $M \sim \Lambda_\QCD$. In this case the complicated potential found on the scalar side dualizes to a function of the form-field field strengths (which are not small compared with $\Lambda_\QCD$ \cite{Dvali:2017mpy, Burgess:2023ifd}). In this case standard semiclassical EFT methods based on derivative expansions can break down on the form-field side.}

This makes duality particularly useful in applications to cosmology, where extremely small scalar masses are on one hand generically required in order for scalars to play an interesting role, but on the other hand are often frowned upon as being technically unnatural. The information that a low-energy scalar field starts life at higher energies in dual form has implications for the size of the couplings to be expected at low energies (see \cite{Reece:2025thc} for a detailed discussion of couplings of axions coming from compactifications of extra dimensions). 

It is indeed true that UV completions can prefer one side of the duality over the other. This can be particularly pertinent for axion physics to the extent that string theory is used to motivate the generic appearance of Goldstone fields at low energies (see {\it e.g.}~\cite{Arvanitaki:2009fg,Cicoli:2012sz} and \cite{Cicoli:2023opf} for a general review) because the axion that is dual to a 4D Kalb Ramond field is often the most robust axion to survive to very low energies in this kind of setup \cite{EarlyStringCompactifications, Cicoli:2012sz,Cicoli:2023opf}.

A related difference between scalars that are duals and generic scalars is the way they couple to other types of matter. For instance an axion $\axion$ is often assumed to couple to matter through a derivative coupling like 
\be \label{AxionCurrent}
    -\frac{\cL_{\rm ax}}{\sqrt{-g}}  = \tfrac12 \, f^2 \partial_\mu \axion \, \partial^\mu \axion +  J^\mu \partial_\axion \,, 
\ee
where $J^\mu$ is some sort of matter current. This kind of coupling also exists on the form side of the duality, with 
\be \label{FormCurrent}
   - \frac{\cL_d}{\sqrt{-g}} = \frac{1}{2\cdot 3!} \, G^{\mu\nu\lambda} G_{\mu\nu\lambda} + \frac{1}{3!} \, \epsilon^{\mu\nu\lambda\rho} J_\mu G_{\nu\lambda\rho}
\ee
where $G = \exd B$ is the two-form's field strength. However the more precise statement of the duality between these interactions is that the interaction \pref{FormCurrent} is actually dual to the lagrangian
\be \label{AxionCurrent2}
    -\frac{\tilde \cL_{\rm ax}}{\sqrt{-g}}  = \tfrac12 \, f^2 (\partial_\mu \axion + J_\mu) ( \partial^\mu \axion +  J^\mu ) \,, 
\ee
which differs from \pref{AxionCurrent} by the addition of a $J^\mu J_\mu$ contact interaction.\footnote{One could instead start from \pref{AxionCurrent} in which case one finds the dual differs from \pref{FormCurrent} by a $J^\mu J_\mu$ term.} More generally, the differences in tensor structure and dimensions of the fields $\axion$ and $G_{\mu\nu\lambda}$ give a different list when writing down the most general couplings to matter that are possible in the low-energy limit. 

A final reason where dual formulations can matter is the case of 3-form gauge potentials (in 4D). In this case the dual does not involve a propagating field and so one might ask: why not just integrate it out once and for all and never discuss the 3-form field? There is more than one answer to this question, but an important one is the role 3-form fields play in conveying to the low-energy world how the UV theory responds to topologically nontrivial situations. They are also useful when discussing scalar mass generation as a Higgs mechanism, since this is accomplished on the dual side by having a 2-form potential eat a 3-form potential.
 
All told, there are three ways that duals can differ in their utility:
\begin{itemize}
\item Weak-strong duality: in which weak dimensionless couplings get mapped to strong couplings. In this case classical solutions still map into one another but the semiclassical expansion need not be good approximation on both sides of the duality. 
\item Topology and non-propagating fields: in which form fields (in 4D especially 3-form fields) bring the news about potentially interesting topology to the low-energy theory.  
\item
Derivative/non-derivative interchange: in which derivative interactions map under duality onto zero-derivative interactions and so risk destablizing the underlying low-energy expansion on which any semiclassical analysis ultimately depends. This is the least well-known of the three, and has implications for the types of matter couplings that arise at low energies and on the size to be expected for the size of interactions within the scalar potential. Besides applying to dualities involving massive scalars and form fields this class of dualities also applies to $F(R)$ gravities \cite{Buchdahl:1970ldb}, which under some circumstances are known to be classically equivalent to a class of scalar-tensor theories. 
\end{itemize}

When scalars arising from duality differ from run-of-the-mill scalars there can be insights to be gained by studying system behaviour on the dual form-field side, where the form-field symmetries play an important role. This might help identify when the mathematical tools devoted to systematizing these generalized symmetries \cite{Gaiotto:2014kfa} might become useful in practice (for reviews see \cite{GenSimRev}).

We organize our findings as follows. \S\ref{sec:DualAxions} starts by reviewing the duality transformations that map massless (and massive) scalars onto 2-form (and 3-form) potentials in 4D spacetimes. This includes a discussion of how matter interactions map under duality and explores some preliminary implications of the $J^\mu J_\mu$ interaction.  \S\ref{ssec:NaturalnessDual} summarizes the power-counting of these theories and how in particular the low-energy expansion is what underlies semiclassical methods in theories with gravity, and how zero-derivative interactions can pose a threat to this expansion. The consistency of the low-energy limit on both sides of the duality is discussed, including how the interactions in the scalar potential can be suppressed enough to make consistency possible. This section closes with a brief discussion of how Starobinsky inflation and $f(R)$ gravity provide a different kind of example of duals with derivative/non-derivative interchange. \S\ref{ssec:NonPropForms} then focusses on 3-form gauge potentials, first describing their appearance in higher dimensional models which illustrate how they bring the news about topological information in the higher dimensions to the low-energy theory. They also  provide explicit examples of how they introduce quadratic structures into the scalar potential that are reminiscent of auxiliary fields in supergravity (not coincidentally, as it turns out).  The section closes by summarizing how 4-form fields arise in some formulations of unimodular gravity and how extra-dimensional models differ from these in ways that help ameliorate their problems. The possibility of vacuum energy relaxation that these models introduce also raise another way in which generic scalars and those that arise as duals can differ: the potentials for generic scalars tend to be supressed along with the rest of the vacuum energy, while dual scalars need not be similarly suppressed. Conclusions and future directions are summarized in \S\ref{sec:Discussion}.
 
\section{Duality for Massless and Massive Forms} 
\label{sec:DualAxions}

This section starts with a review of the duality construction relating massless and massive scalars to form fields in 4 spacetime dimensions, and then passes to a discussion of the significance of an underappreciated contact interaction that is often ignored though it must appear on one side of the duality or the other. 

\subsection{Duality for massless and massive scalars}
\label{ssec:MasslessMassiveD}

We start with a review of form-scalar duality for massless and massive scalar fields in four spacetime dimensions (for a review see \cite{Polchinski:2014mva, Quevedo:1997jb}). The restriction to 4D is both for simplicity and because this is the case of practical interest. Similar arguments go through for other dimensions but are not needed here (with the exception of a 6D example used in passing in \S\ref{sssec:SalamSezgin}).

\subsubsection{Massless case}
\label{ssec:MasslessD}

Consider the following path integral over fields in four spacetime dimensions:
\be
  \Xi[J ,Z] =  \int \cD B \; e^{iS_1[B]}
\ee
where $S_1 = \int \exd^4 x \; \cL_1$ with $\cL_1$ given by 
\be \label{KRForm1}
   \cL_1(B) =  - \frac{Z}{2 \cdot 3!} \, G_{\mu\nu\lambda} G^{\mu\nu\lambda}  -   \frac{1}{3!} \, \epsilon^{\mu\nu\lambda\rho} G_{\mu\nu\lambda} J _{\rho}   \,.
\ee
Here the 3-form $G = \exd B$ is the exterior derivative of the 2-form field $B_{\mu\nu}$ and both the current $J _\mu$ and the pre-factor $Z$ can be functions of other fields. $B$ is only defined up to the gauge redundancy $B \to B + \exd  \lambda$ for an arbitrary 1-form $\lambda_\mu$. Notice also that the Bianchi identity $\exd G = 0$ implies the lagrangian transforms into a total derivative if $J $ is redefined by $J  \to J  + \exd \omega$, for some scalar field $\omega$, so $J $ effectively behaves like a gauge potential. 

$Z$ is dimensionless if $B$ has canonical dimension of mass (in particular within extra-dimensional models one often finds $Z = e^{-\beta \varphi}$ for a dilaton field $\varphi$). If, on the other hand, $B$ has dimension (mass)${}^{2}$ -- as would be true if it were to satisfy a flux quantization condition wherein $\int _\ssM B$ over a 2-dimensional manifold $M$ is proportional to an integer -- then $Z$ has dimension (mass)${}^{-2}$ and so can be written $Z = 1/f^{2}$ for some mass scale $f$. 

The duality starts by trading the integration over $B$ for an integral over $G$ subject to the Bianchi identity $\exd G = 0$, imposed as a constraint using a scalar Lagrange-multiplier field $\axion$,  
\be
  \Xi[J ,f] =  \int \cD G  \, \cD \axion \; e^{iS_0[G, \mfa]}
\ee
where $S_0 = \int \exd^4 x \; \cL_0$ with
\be \label{MasterDual}
   \cL_0(G,\axion) =  - \frac{Z}{2\cdot 3!} \, G_{\mu\nu\lambda} G^{\mu\nu\lambda} - \frac{1}{3!} \axion \, \epsilon^{\mu\nu\lambda\rho} \partial_{\mu} G_{\nu\lambda\rho}  - \frac{1}{3!} \, \epsilon^{\mu\nu\lambda\rho} G_{\mu\nu\lambda} J _{\rho} \,,
\ee
Integrating out $\axion$ imposes the Bianchi identity $\exd G = 0$ and allows the integral over $G$ to be replaced with the integral over $B$, leading back to \pref{KRForm1}. 

The dual is obtained by integrating out $G_{\mu\nu\lambda}$ so that it is the field $\axion$ that remains.  This can be done explicitly here because the required integral is gaussian. Notice the emergence of a shift symmetry $\axion \to \axion \;+ $ constant under which $\cL_0$ transforms into a total derivative. The $G$ integration is gaussian with saddle point $G_{\mu\nu\lambda} = \cG_{\mu\nu\lambda}$ where
\be \label{GSaddle}
    \cG_{\mu\nu\lambda} = - Z^{-1} \epsilon_{\mu\nu\lambda\rho} \Bigl( \partial^\rho \axion + J ^\rho \Bigr) \,,
\ee
and gives the dual lagrangian
\be \label{Dualmassless}
   \cL_2(\axion) 
   =  - \tfrac{1}{2} Z^{-1}(\partial_\mu \axion + J _\mu)( \partial^\mu \axion +  J ^\mu) + \cL_{\rm st}\,,
\ee
so if $J$ were a fundamental field it would gauge the axion shift symmetry. In \pref{Dualmassless} $\cL_{\rm st}$ denotes the surface term that arises due to the integration by parts required to obtain the saddle-point condition:
\bea \label{stform}
   \cL_{\rm st} &=&  - \frac{1}{3!} \partial_{\mu} \Bigl( \axion \, \epsilon^{\mu\nu\lambda\rho}  \cG_{\nu\lambda\rho} \Bigr) =   \frac{1}{3!} \partial_{\mu} \Bigl[ Z^{-1}  \axion \, \epsilon^{\mu\nu\lambda\rho}  \epsilon_{\nu\lambda\rho\sigma}(\partial^\sigma \axion + J^\sigma) \Bigr] \nn\\
   &=&  \partial_\mu \Bigl[ Z^{-1}  \axion \,  (\partial^\mu \axion + J^\mu) \Bigr]  \,.
\eea 
It is convenient to define the (possibly field-dependent) quantity $f$ by $Z = 1/f^2$ so that the scalar kinetic term has the form $\frac12 f^2 (\partial \axion)^2$. If $B$ has canonical dimension (mass) in fundamental units then so does $\axion$ and then $Z$ and $f$ are both dimensionless. But if instead $B$ has dimension (mass)${}^{2}$  then $\axion$ is dimensionless and $f$ has dimension mass.  

\subsubsection{Massive case}
\label{sssec:MassiveCase}

We next broaden the context by describing how duality works for massive scalars \cite{Quevedo:1996uu}. Consider therefore the path integral
\be
  \Xi[J ,\cJ, \cZ,f] = \int \cD C \, \cD B  \; e^{iS_1(C,B)}
\ee
with 3-form gauge potential $C$ and 2-form gauge potential $B$ and action $S_1 = \int \exd^4 x \; \cL_1$ with
\be \label{L1hatG}
   \cL_1(C,B) = - \frac{\cZ^2}{2\cdot 4!} H_{\mu\nu\lambda\rho} H^{\mu\nu\lambda\rho} - \frac{1}{2\cdot 3! f^2} \widehat G_{\mu\nu\lambda} \widehat G^{\mu\nu\lambda}  - \frac{1}{3!} \, \epsilon^{\mu\nu\lambda\rho}  \widehat G_{\mu\nu\lambda} J _\rho   - \frac{1}{4!} \, \cJ \, \epsilon^{\mu\nu\lambda\rho}  H_{\mu\nu\lambda\rho}   \,,
\ee
where $H = \exd C$ satisfies the usual Bianchi identity $\exd H = 0$ but now $\widehat G$ instead satisfies the modified Bianchi identity $\exd \widehat G = m H$ where $m$ is a parameter with dimension mass. This has as its local solution $\widehat G = G + m C$ (where $G = \exd B$ as before). $J_\mu$ and $\cJ$ can be external currents or functions of other fields.

This lagrangian has the two independent gauge symmetries $B \to B + \exd \lambda$ and  $(C, B) \to (C + \exd \Lambda , B \to B - m\, \Lambda)$ for an arbitrary 1-form $\lambda$ and arbitrary 2-form $\Lambda$. Notice that when $m \neq 0$ we can choose $\Lambda$ to fix a gauge $B = 0$, in which case the lagrangian looks like the lagrangian for a massive 3-form potential
\be \label{UnitaryGauge}
   \cL_1(C,B) = - \frac{\cZ^2}{2\cdot 4!} H_{\mu\nu\lambda\rho} H^{\mu\nu\lambda\rho} - \frac{m^2}{2\cdot 3! f^2} C_{\mu\nu\lambda}  C^{\mu\nu\lambda}   - \frac{m}{3!} \, \epsilon^{\mu\nu\lambda\rho}  C_{\mu\nu\lambda}  J _\rho   - \frac{1}{4!} \, \cJ \, \epsilon^{\mu\nu\lambda\rho}  H_{\mu\nu\lambda\rho}\,.
\ee

The dual is obtained by starting with \pref{L1hatG} but trading the integral over $B$ for an integral over $\widehat G$, 
\bea \label{L0Ghat}
   \cL_0(C,G,\axion) &=& - \frac{\cZ^2}{2\cdot 4!} H_{\mu\nu\lambda\rho} H^{\mu\nu\lambda\rho} - \frac{1}{2\cdot 3! f^2} \widehat G_{\mu\nu\lambda} \widehat G^{\mu\nu\lambda}  - \frac{1}{3!}\, \axion \, \epsilon^{\mu\nu\lambda\rho} \Bigl( \partial_\mu \widehat G_{\nu\lambda\rho}  - \tfrac14 m H_{\mu\nu\lambda \rho} \Bigr) \nn\\
   && \qquad\qquad\qquad - \frac{1}{3!} \, \epsilon^{\mu\nu\lambda\rho} \widehat G_{\mu\nu\lambda} J _\rho  - \frac{1}{4!} \, \cJ \, \epsilon^{\mu\nu\lambda\rho}  H_{\mu\nu\lambda\rho} \,.
\eea
Integrating out $\axion$ gives back \pref{L1hatG} but to obtain the dual we instead integrate over $\widehat G = G + m C$ using the saddle point $G = \cG$ where
\be \label{GSaddlem}
    \cG_{\mu\nu\lambda} = - m C_{\mu\nu\lambda} - f^2 \epsilon_{\mu\nu\lambda\rho} \Bigl( \partial^\rho \axion + J ^\rho \Bigr) \,.
\ee
This gives the lagrangian 
\be \label{Haform}
   \cL_2(C,\axion) 
   = - \frac{\cZ^2}{2\cdot 4!} H_{\mu\nu\lambda\rho} H^{\mu\nu\lambda\rho}  + \frac{1}{4!}\,  (m \axion - \cJ) \, \epsilon^{\mu\nu\lambda\rho} H_{\mu\nu\lambda\rho}-  \frac{f^2}2   \Bigl( \partial_\mu \axion + J _\mu \Bigr) \Bigl( \partial^\mu \axion + J ^\mu \Bigr) + \cL_{\rm st} \,,
\ee
where $\cL_{\rm st}$ is given by \pref{stform}:
\be \label{st3form}
   \cL_{\rm st} =  - \frac{1}{3!} \partial_\mu \Bigl( \axion \, \epsilon^{\mu\nu\lambda\rho}  \cG_{\nu\lambda\rho} \Bigr) =   \frac{1}{3!} \partial_\mu \Bigl[ f^2  \axion \, \epsilon^{\mu\nu\lambda\rho}  \epsilon_{\nu\lambda\rho\sigma} (\partial^\sigma \axion + J^\sigma) \Bigr] = \partial_\mu \Bigl[ f^2  \axion \,  (\partial^\mu \axion + J^\mu) \Bigr]  \,.
\ee

The final step is to perform the integral over $C_{\mu\nu\lambda}$, which is equivalent\footnote{It is not quite equivalent because integrating over $C$ instead gives the equation $\nabla_\mu (\cZ^2 \cH^{\mu\nu\lambda\rho} - m \axion \epsilon^{\mu\nu\lambda\rho}) = 0$, which on integration allows the solution to involve functions with zero divergence in addition to the one used here. Appendix \ref{App:Scalar-4form} shows why this complication does not affect the arguments we give here.} to simply integrating over $H_{\mu\nu\lambda\rho}$ because the integrability condition for writing $H = \exd C$ is $\exd H = 0$ which is always true in 4D. The saddle point for the $H$ integral occurs for $H_{\mu\nu\lambda\rho} = \cH_{\mu\nu\lambda\rho}$ where
\be \label{HSaddle}
   \cH_{\mu\nu\lambda\rho} =  \frac{1}{ \cZ^2} (m \axion - \cJ) \, \epsilon_{\mu\nu\lambda\rho}
\ee
and so leads to the scalar lagrangian
\be \label{dualscalarmassive}
   \cL_2(\axion)
  = -\frac{f^2}2  \Bigl( \partial_\mu \axion + J _\mu \Bigr) \Bigl( \partial^\mu \axion + J ^\mu \Bigr)  - \frac{1}{2\cZ^2} (m \axion - \cJ)^2  + \cL_{\rm st}\,.
\ee
As advertised, this contains the potential that makes $\axion$ a massive scalar whose mass is
\be \label{axionmass}
   m_a = \frac{m}{\cZ f} \,.
\ee

\subsubsection{Including anomaly terms}

Broadening the form of the Bianchi identity also suggests a generalization. Rather than imposing $\exd \widehat G = m H$ we can instead demand
\be \label{NewConstraint}
    \exd \widehat G = m H + \Omega 
\ee
where $\Omega$ is a gauge-invariant 4-form built from other fields that is locally exact ({\it i.e.}~there exists a 3-form $\omega$ such that $\Omega = \exd \omega$). The local solution to \pref{NewConstraint} would then be 
\be \label{GChern}
   \widehat G = \exd B + m C + \omega \,.
\ee
For example one might have $\Omega \propto \Tr (F \wedge F)$ where $F_{\mu\nu}$ is a matrix-valued gauge field strength, in which case $\omega$ is the corresponding Chern-Simons form. 

Imposing \pref{NewConstraint} using a lagrange multiplier then replaces \pref{L0Ghat} with
\bea \label{L0Ghat2}
   \cL_0(C,G,\axion) &=& - \frac{\cZ^2}{2\cdot 4!} H_{\mu\nu\lambda\rho} H^{\mu\nu\lambda\rho} - \frac{1}{2\cdot 3! f^2} \widehat G_{\mu\nu\lambda} \widehat G^{\mu\nu\lambda}  - \frac{1}{3!} \, \epsilon^{\mu\nu\lambda\rho} \widehat G_{\mu\nu\lambda} J _\rho  - \frac{1}{4!} \, \cJ \, \epsilon^{\mu\nu\lambda\rho}  H_{\mu\nu\lambda\rho}\nn\\
   && \qquad\qquad\qquad  - \frac{1}{3!}\, \axion \, \epsilon^{\mu\nu\lambda\rho} \Bigl( \partial_\mu \widehat G_{\nu\lambda\rho}  - \tfrac14 m H_{\mu\nu\lambda \rho} - \tfrac14 \Omega_{\mu\nu\lambda\rho} \Bigr) \,.
\eea
Integrating out $\widehat G$ using the same saddle point \pref{GSaddlem} then replaces \pref{Haform} with
\bea \label{Haform2}
   \cL_2(C,\axion) &=& - \frac{\cZ^2}{2\cdot 4!} H_{\mu\nu\lambda\rho} H^{\mu\nu\lambda\rho}   -  \frac{f^2}2   \Bigl( \partial_\mu \axion + J _\mu \Bigr) \Bigl( \partial^\mu \axion + J ^\mu \Bigr)  \nn\\
   && \qquad\qquad\qquad  + \frac{1}{4!}\,  \axion \, \epsilon^{\mu\nu\lambda\rho} \Bigl(m  H_{\mu\nu\lambda\rho} +  \Omega_{\mu\nu\lambda\rho} \Bigr)- \frac{1}{4!} \, \cJ \, \epsilon^{\mu\nu\lambda\rho}  H_{\mu\nu\lambda\rho} + \cL_{\rm st}\,,
\eea
and so performing the final integral over $H$ leads to the dual scalar lagrangian
\be \label{L2axionmassive}
   \cL_2(\axion)
  = -\frac{f^2}2  \Bigl( \partial_\mu \axion + J _\mu \Bigr) \Bigl( \partial^\mu \axion + J ^\mu \Bigr)  - \frac{1}{2\cZ^2}(m \axion - \cJ)^2  + \frac{1}{4!}\,  \axion \, \epsilon^{\mu\nu\lambda\rho}  \Omega_{\mu\nu\lambda\rho}  + \cL_{\rm st} \,.
\ee
instead of \pref{dualscalarmassive}. As before $\cL_{\rm st}$ is given by \pref{st3form}. The $\Omega$-dependent term represents a gauge-axion anomaly type coupling and the generalization argument shown here shows how these are dualized as the inclusion of the Chern-Simons form in \pref{GChern}.

For later purposes what matters is the mass term proportional to $(m \axion - \cJ)^2$ on the scalar side starts life on the form side as the 4-form `kinetic' term $H_{\mu\nu\lambda\rho} H^{\mu\nu\lambda\rho}$, as can be seen from the saddle-point condition \pref{HSaddle}. This shows how the duality maps nonderivative terms in the action on the scalar side onto derivative interactions on the form side.\footnote{One might be tempted to discount the derivative interactions since the $H_{\mu\nu\lambda\rho}$ `only' describes the kinetic energy of a non-propagating field. The same could also be said for some of the components of the electromagnetic field and to Fadeev-Popov-DeWitt ghosts in a covariant gauge (which is the gauge in which loop calculations are usually actually performed) for which power-counting are usefully applied. \S\ref{ssec:NonPropForms} below provides other arguments on the utility of keeping non-propagating fields explicitly in semiclassical calculations.}

\subsection{The general case}
The results of the previous sections can be straightforwardly generalized to the $D$-dimensional case of massless $p$ forms dual to massless $D-p-2$ forms and massive $p$ forms dual to massive $q=D-p-1$ forms. 

For the massive case this means we start with the master Lagrangian for fields $C_p, G_p, A_q$ (with $A_q$ generalizing the scalar axion field):
\bea \label{L0GhatGen}
   \cL_0[C_p,G_p,A_q]=  &-& \frac{\cZ^2}{2\cdot (p+1)!} H_{\mu_1\cdots \mu_{p+1}} H^{\mu_1\cdots \mu_{p+1}} - \frac{1}{2\cdot p! f^2} \widehat G_{\mu_1\cdots \mu_p} \widehat G^{\mu_1\cdots \mu_p}\nn\\  &-& \frac{1}{p!} \, \epsilon^{\mu_1\cdots \mu_D} \Bigl( \partial_{\mu_1} \widehat G_{\mu_2\cdots \mu_{p+1}}  - \frac{1}{p+1} m H_{\mu_1\cdots \mu_{p+1}}\Bigr) A_{\mu_{p+2}\cdots \mu_D}\nn\\
   &-&  \frac{1}{p!} \, \epsilon^{\mu_1\cdots \mu_D} \widehat G_{\mu_1\cdots \mu_p} J _{\mu_{p+1}\cdots \mu_D}  - \frac{1}{(p+1)!} \,  \epsilon^{\mu_1\cdots \mu_D}  H_{\mu_1\cdots \mu_{p+1}}\, \cJ_{\mu_{p+2}\cdots \mu_D} \,.
\eea
Here again $H_{p+1}=\exd C_p$ and the currents are general forms  $J_{q+1}, \mc J_{p+1}$.  Following the same procedure as before, we can get two dual Lagrangians. The first of these is
\bea \label{UnitaryGaugeGen}
   \cL_1[C_p] = &-& \frac{\cZ^2}{2\cdot (p+1)!} H_{\mu_1\cdots\mu_{p+1}} ^2 - \frac{m^2}{2\cdot p! f^2} C_{\mu_1\cdots \mu_p}^2     \nn \\ &-& \frac{1}{p!} \, \epsilon^{\mu_1\cdots \mu_D}\left( m\, C_{\mu_1\cdots \mu_p} J _{\mu_{p+1}\cdots \mu_D}   - \frac{1}{(p+1)}   H_{\mu_1\cdots\mu_{p+1}}\, \cJ_{\mu_{p+2}\cdots\mu_D}\right) \,,
\eea
describing the Lagrangian for a massive $p$-form field $C_{\mu_1\cdots \mu_p}$, where again $H=\exd C$. The dual Lagrangian for the $q$-form field $A_{\mu_1\cdots\mu_{q}}$ with $q=D-p-1$ is:
\be \label{dualscalarmassiveGen}
   \cL_2[A_q]
  = -\frac{f^2}2  \Bigl( \partial_{\mu_1} A_{\mu_2\cdots \mu_{q+1}}+ J _{\mu_1\cdots \mu_{q+1}} \Bigr)^2   - \frac{1}{2\cZ^2} \left(m A_{\mu_1\cdots \mu_q} - \cJ_{\mu_1\cdots \mu_q}\right)^2  + \cL_{\rm st}\,.
\ee

We now make more explicit that dual theories map any complicated non-derivative couplings to equally complicated derivative couplings. Formally, a general way to write these dualities is in terms of generalized Fourier transforms as follows (here we generalize a procedure discussed in \cite{Kawai:1980qq, Diamantini:1996vf}). Consider a $p$ form $C_p$ and a $q=D-p-1$ form $ A_{q}$. Consider the following path integral  for the fields $C_p$ and ${ A}_q$ coupled to currents $J_{q+1}$ and ${\mathcal J}_{p+1}$ respectively:
\be
{\Xi}\left[J_{p-1}, {\mc J}_{q-1}\right]=\int {\mathcal D} C_{p}\, {\mathcal D}{A}_{q}\, e^{i\int d^\ssD x \, {\mathcal L}\left(C,{A}, J, {\mathcal J} \right)}
\ee
with 
\bea
{\mathcal L}_0\left[C_p,A_q\right] &=& \epsilon^{\mu_1\cdots\mu_D}C_{\mu_1\cdots \mu_p}\, \partial_{\mu_{p+1}}{A}_{\mu_{p+2}\cdots {\mu_D}}+F\left(C_p\right)+{\tilde G}\left(\tilde A_q\right)\nn \\
&+&   \epsilon^{\mu_1\cdots\mu_D}C_{\mu_1\cdots \mu_p}\,J_{\mu_{p+1}\cdots \mu_D}+ \epsilon^{\mu_1\cdots\mu_D}{A}_{\mu_1\cdots \mu_q}\mc J_{\mu_{q+1}\cdots \mu_D}
\eea
where $F(C_p)$ and ${\tilde G}({A}_q)$ are arbitrary functions of their arguments.

We can then follow the standard duality procedure. Integration  over $A_q$ (after integrating by parts the first term) corresponds to a Fourier transformation (with dual variables $A_q$ and $ \exd{C}_p$ ) which formally leads to the Lagrangian 
\be\label{Ltwo}
{\mathcal L}_1\left[C_p\right]= G\left(\exd C_p-{\mathcal J}_{p+1}\right)+F\left( C_p\right)+\epsilon^{\mu_1\cdots\mu_D}\, C_{\mu_1\cdots \mu_p}\,J_{\mu_{p+1}\cdots \mu_D}
\, +\, {\mathcal L}_{\rm{st}}
\ee
where now $e^G$ is the Fourier transform of $e^{\tilde G}$. Similarly, integration over ${{C}}_p$  leads to the dual theory with Lagrangian:
\be\label{Lone}
{\mathcal L}_2\left[A_q\right]={\tilde F}\left(\exd{A}_q-J_{q+1}\right)+{\tilde G}({A}_{q})+ \epsilon^{\mu_1\cdots\mu_D}{A}_{\mu_1\cdots \mu_q}\mc J_{\mu_{q+1}\cdots \mu_D}
\ee
where $e^{i\tilde{F}}$ is the Fourier transform of $e^{iF}$.

From here we can derive all the $p$-forms dualities as special cases:
\begin{enumerate}
\item Let us take $\tilde G=\mc J=0$ and $F$  a quadratic function. We know that for dual coordinates $X,P$, the Fourier transform of $e^{i\alpha X}$ is $\delta(P-\alpha)$. This implies $e^{G}=\delta(\exd C_p)$, fixing $\exd C_p=0$. Then locally $C_p=\exd B_{p-1}$ and equation \ref{Ltwo} gives the Lagrangian for a massless $p-1$ form $B_{p-1}$ with $F(\exd B_{p-1})$ providing the (canonical) kinetic term. On the other hand, equation \ref{Lone} gives $\mc L_1\propto \left(\exd A-J\right)^2$ (since the Fourier transform of a Gaussian is another Gaussian) which describes a massless $q$-form $A_q$ ($q=D-p-1$) coupled to the current $J_{q+1}$. Then we recover the standard duality among massless forms  $B_{p-1}$ and $A_q$, generalizing the results of the previous sections.
\item If instead $F$ and $J_{q+1}$ vanish and $\tilde G$ is quadratic, we recover the same massless duality but among $q-1$ and $p$ forms. 
\item If  ${\tilde G}$ and $F$ are quadratic we reproduce the dualities among {\it massive} fields $C_p$ and $A_q$ that we have been discussing (using again that Fourier transforms of Gaussians are Gaussians).
\item In general here we can see that a potentially complicated function $F(C_p)$ (corresponding to non derivative couplings for the field $C_p$) can generate an equally complicated function $\tilde F$ of the derivatives of $A_p$. The presence of the implied complicated derivative interactions limits the ability to evaluate the resulting path integrals semiclassically, limiting the information content that can in practice be extracted from the form-field side.
\end{enumerate}

The last of these points implies in particular that if one side of the duality is deformed using non-derivative interactions then it is unlikely that the deformation is also derivative-free on the other side. If one compares the two dual theories by adding non-derivative interactions to {\it both} of them -- as was done for instance in \cite{Hell:2021wzm} -- one expects to (and does) find inconsistencies that break the equivalence between the two systems.

Notice that there is no claim that non quadratic potentials, like $\cos\mfa$ are impossible to achieve in systems that admit dual descriptions (see {\it e.g.} \cite{Diamantini:1996vf, Dvali:2005an}). But when these duals exist they include higher-derivative interactions that undermine the utility of semiclassical EFT methods on the dual side.\footnote{In many ways the situation is similar to what happens when one applies the bosonization transformation to fermions in dimensions higher than two. Although a well-defined prescription still exists for doing so in higher dimensions the corresponding path integrals are difficult to evaluate reliably and so the dual theory seems not that informative \cite{Burgess:1994tm}.} So although the above Fourier transform approach makes explicit the connection between non-derivative and derivative couplings in the two dual theories, in practice to have an EFT under control on both sides requires formulating well-defined low-energy expansions for each of the dual theories (the conditions for which are described in more detail in \S\ref{ssec:NaturalnessDual}).

\subsection{Contact interactions and frustration}

Before exploring the conditions for semiclassical expansion in more detail we first pause to explore the significance of the contact $J_\mu J^\mu$ interaction seen on the scalar side in \pref{L2axionmassive}. To this end this section explores some of the simple solutions in the classical field equations for the scalar and its dual in the presence of matter. 

To do so we return to the action in the pre-gauge-fixed form \pref{L1hatG}, repeated here for convenience (now with metrics inserted):
\be \label{L1hatG2}
   \cL(C,B) = - \sqrt{-g} \left[ \frac{\cZ^2}{2\cdot 4!} H_{\mu\nu\lambda\rho} H^{\mu\nu\lambda\rho} + \frac{1}{2\cdot 3! f^2} \widehat G_{\mu\nu\lambda} \widehat G^{\mu\nu\lambda} + \frac{1}{3!} \, \epsilon^{\mu\nu\lambda\rho}  \widehat G_{\mu\nu\lambda} J _\rho  + \frac{1}{4!} \, \cJ \, \epsilon^{\mu\nu\lambda\rho}  H_{\mu\nu\lambda\rho} \right]   \,,
\ee
where $\epsilon_{\mu\nu\lambda\rho} = \pm \sqrt{-g}$ (rather than its Minkowski-space value $\pm 1$). Notice in particular that this means only the first two terms in this lagrangian are metric dependent. The field equations for the field $C_{\mu\nu\lambda}$ coming from the lagrangian \pref{L1hatG2} are
\bea \label{CEOM}
   0 &=&  \frac{1}{\sqrt{-g}} \partial_\mu \Bigl[ \sqrt{-g} \Bigl( \cZ^2 H^{\mu\nu\lambda\rho} + \cJ \epsilon^{\mu\nu\lambda\rho} \Bigr)\Bigr] - m \left(\frac{1}{f^2} \widehat G^{\nu\lambda\rho} + \epsilon^{\nu\lambda\rho\mu} J_\mu\right)  \nn\\
   &=&  \nabla_\mu \Bigl( \cZ^2 H^{\mu\nu\lambda\rho} + \cJ \epsilon^{\mu\nu\lambda\rho} \Bigr) - m \left(\frac{1}{f^2} \widehat G^{\nu\lambda\rho} + \epsilon^{\nu\lambda\rho\mu} J_\mu\right)  \,,
\eea
and the field equations coming from varying $B_{\mu\nu}$ are similarly
\be \label{BEOM}
    \partial_\mu \left[ \sqrt{-g} \left( \frac{1}{f^2} \widehat G^{\mu\nu\lambda} + \epsilon^{\mu\nu\lambda\rho} J_\rho \right)\right] =   \sqrt{-g\;}\nabla_\mu  \left( \frac{1}{f^2} \widehat G^{\mu\nu\lambda} + \epsilon^{\mu\nu\lambda\rho} J_\rho \right) = 0\,.
\ee
When $m \neq 0$ eq.~\pref{BEOM} can also be obtained by taking the divergence of \pref{CEOM} and using $\nabla_\nu \nabla_\mu[ f \epsilon^{\mu\nu\lambda\rho}] = 0$ for any scalar $f$. 

The stress energy appearing in Einstein equations implied by \pref{L1hatG2} is
\bea
   T^{\mu\nu} = \frac{2}{\sqrt{-g}} \, \frac{\delta S}{\delta g_{\mu\nu}} 
   &=&  \frac{\cZ^2}{ 3!} {H^\mu}_{\beta\lambda\rho} H^{\nu\beta\lambda\rho} + \frac{1}{ 2! f^2} \widehat {G^\mu}_{\beta\lambda} \widehat G^{\nu\beta\lambda}  \nn\\
  &&\qquad  -  \left[ \frac{\cZ^2}{2\cdot 4!} H_{\alpha\beta\lambda\rho} H^{\alpha\beta\lambda\rho} + \frac{1}{2\cdot 3! f^2} \widehat G_{\alpha\beta\lambda} \widehat G^{\alpha\beta\lambda} \right] g^{\mu\nu} \,.
\eea
Specializing to a static metric of the form
\be
    \exd s^2 = - \exd t^2 + \gamma_{ij}(x) \, \exd x^i \, \exd x^j \,,
\ee
gives the energy density
\be \label{formenergy}
 \rho = T^{00} =  \frac{\cZ^2}{ 3!} \Bigl[ {H^0}_{\beta\lambda\rho} H^{0\beta\lambda\rho}   +   \tfrac14 H_{\alpha\beta\lambda\rho} H^{\alpha\beta\lambda\rho}\Bigr] + \frac{1}{ 2! f^2} \Bigl[ \widehat {G^0}_{\beta\lambda} \widehat G^{0\beta\lambda}  + \tfrac16 \widehat G_{\alpha\beta\lambda} \widehat G^{\alpha\beta\lambda} \Bigr] \,,
\ee
while the spatial stress is
\be \label{formstress}
   T^{ij}  =  \frac{\cZ^2}{ 3!}  \Bigl[ {H^i}_{\beta\lambda\rho} H^{j\beta\lambda\rho}   -   \tfrac14 \gamma^{ij} H_{\alpha\beta\lambda\rho} H^{\alpha\beta\lambda\rho}\Bigr] + \frac{1}{ 2! f^2} \Bigl[ \widehat {G^i}_{\beta\lambda} \widehat G^{j\beta\lambda}  - \tfrac16 \gamma^{ij} \widehat G_{\alpha\beta\lambda} \widehat G^{\alpha\beta\lambda} \Bigr]   \,.
\ee
To the extent that dualization maps $H \leftrightarrow m\axion$ and $\widehat G \leftrightarrow \partial \axion$ the $H^2$ terms can be regarded as the potential contributions to the energy and pressure while the $\widehat G^2$ terms give gradient contributions.

Next specialize these equations to unitary gauge, for which $B_{\mu\nu} = 0$ and so $\widehat G_{\mu\nu\lambda} = m C_{\mu\nu\lambda}$. The field equations for $C_{\mu\nu\lambda}$ then simplify to
\be
    \nabla_\mu \Bigl( \cZ^2 H^{\mu\nu\lambda\rho} + \cJ \epsilon^{\mu\nu\lambda\rho} \Bigr)  - m \left( \frac{m}{f^2}C^{\nu\lambda\rho} + \epsilon^{\nu\lambda\rho\mu} J_\mu\right) = 0 \,,
\ee
and (when $m \neq 0$) the $B_{\mu\nu}$ equation can be found by differentiating this:
\be
  \nabla_\nu \left( \frac{m}{f^2}C^{\nu\lambda\rho} + \epsilon^{\nu\lambda\rho\mu} J_\mu\right) = 0  \,.
\ee

In the special case where $\cZ$, $\cJ$ and $f$ are constants and using $m_a = m/(f\cZ)$ the field equations become
\be \label{divC}
   \nabla_\mu   H^{\mu\nu\lambda\rho} - m^2_a C^{\nu\lambda\rho} = \frac{m}{\cZ^2} \epsilon^{\nu\lambda\rho\mu} J_\mu \quad\hbox{and} \quad  - m \nabla_\nu C^{\nu\lambda\rho} = f^2 \epsilon^{\nu\lambda\rho\mu} \partial_\nu J_\mu \,.
\ee
Notice in particular the current only acts as a source in the second of these equations if its curl is nonzero. If $J^\mu$ satisfies $\exd J = 0$ (such as if $J$ vanishes or is a constant) then $\nabla_\mu C^{\mu\nu\lambda} = 0$. Then the first equation becomes $(\Box - m^2_a) C^{\mu\nu\lambda} = S^{\mu\nu\lambda}$, describing waves for a single spin state of mass $m_a$ coupled to the source $S^{\mu\nu\lambda} := m \epsilon^{\mu\nu\lambda\rho} J_\rho/\cZ^2$. 

\subsubsection{Minimal static solutions: form side}

Consider next (non-wave) solutions that arise in the immediate environment of simple weakly gravitating sources. Suppose for instance that $J_k = 0$ and $J_0$ is nonzero but time-independent, where we split up time and space directions, $x^\mu \to \{x^0, x^i \}$. This allows \pref{CEOM} to be written (in the flat limit)
\be \label{FlatspaceBeq2}
  \partial_t \left( \frac{m}{f^2}C^{0 ij}  \right) + \partial_k \left( \frac{m}{f^2}C^{ ijk} + \epsilon^{ ijk} J^0 \right)= 0  \quad\hbox{and} \quad
   \partial_j \left( \frac{m}{f^2}C^{0ij}  \right) = 0  \,.
\ee
while \pref{BEOM} becomes the pair
\be \label{FlatspaceCeq1}
     \partial_t   \Bigl( \cZ^2 H^{0ijk} + \cJ \epsilon^{ijk} \Bigr)  - \frac{m^2}{f^2}\, C^{ijk} = m \epsilon^{ijk} J^0 \,,
\ee
and
\be \label{FlatspaceCeq2}
     \partial_j   \Bigl( \cZ^2 H^{0ijk} + \cJ \epsilon^{ijk} \Bigr)  - \frac{m^2}{f^2}\, C^{0ik} = m \epsilon^{ikj} J_j = 0\,.
\ee

We seek a static solution that everywhere satisfies
\be \label{CijkForm}
m C^{ijk} = - f^2 \epsilon^{ijk} J^0 \,,
\ee
where $J^0$ can vary in space but not time. Then \pref{FlatspaceBeq2} becomes
\be \label{FlatspaceBeq2a}
  \partial_t \left( \frac{m}{f^2}C^{0 ij}  \right)  =  
   \partial_j \left( \frac{m}{f^2}C^{0ij}  \right) = 0  \,.
\ee
If we also assume $m$ and $f$ are constants then $\dot C^{0ij} = 0$ (where overdots denote differentiation with respect to $t$) and $\partial_j C^{0ij} = 0$ and so $C^{0ij} = \epsilon^{ijk} \partial_k \mfC$ for some time-independent $\mfC$. The field-strength derivatives then become $\partial_t H^{0ijk} = -   \ddot C^{ijk}$ and $\partial_j H^{0ijk} =    \partial_j \dot C^{ijk}  + \epsilon^{ikl} \partial_j \partial^j \partial_l \mfC$.
 
Using \pref{CijkForm} and the time-independence of $J^0$ implies $\dot C^{ijk} = 0$ and so $\partial_t H^{0ijk} = 0$ and $\partial_j H^{0ijk} =  \epsilon^{ikl} \partial_j \partial^j \partial_l \mfC$. Assuming $\cJ$and $\cZ$ are also constants, the remaining field equations boil down to
\be \label{FlatspaceCeq2a}
      \Bigl( \partial_j \partial^j  -  m_a^2 \Bigr) \partial_l \mfC =  0\,,
\ee
where $m_a = m/(\cZ f)$ (as before). This last equation contains the wave information (and shows the equivalence to the scalar waves in the dual theory).  Because $J_j = 0$ and since our interest here is not in waves we can choose $\mfC = 0$ consistent with the equations of motion, leaving \pref{CijkForm} as the only condition. 

Since $J^0$ is assumed time-independent the field strengths to which we are led are
\be \label{FinalStaticForm}
   \widehat \cG_{ij0} = m \cC_{ij0} = 0 \,, \qquad \widehat \cG_{ijk} = m \cC_{ijk} =  -f^2 \epsilon_{ijk} J^0 \,,
   \quad \hbox{and} \quad
   \cH = \exd \cC = 0 \,,
\ee
where the vanishing of $\cH$ follows from the time-independence of $f^2 J^0$. 

\subsubsection{Minimal static solutions: scalar side}

Comparing expressions \pref{FinalStaticForm} with the saddle points \pref{GSaddle} and \pref{HSaddle}, which say
\be
  \epsilon^{\mu\nu\lambda\rho} \cH_{\mu\nu\lambda\rho} = - 4! ( m \axion - \cJ) \quad \hbox{and} \quad
  \widehat \cG_{\mu\nu\lambda} =  -f^2 \epsilon_{\mu\nu\lambda\rho} (\partial^\rho \axion + J^\rho) \,, 
\ee
shows that the dual $\axion$ configuration satisfies $m\axion = \cJ$, which by assumption is constant so $\partial_\mu \axion = 0$. 

Does this mean that the solutions are trivial? To check, we evaluate the solution's stress energy by specializing \pref{formenergy} and \pref{formstress} to the case where $B_{\mu\nu} = C_{0ij} = 0$ and $C_{ijk}$ is time-independent. With these choices we have $H_{\mu\nu\lambda\rho} = \widehat G_{0ij}  = 0$ and $\widehat G_{ijk} = m C_{ijk}$ and so
\be \label{BCT00soln}
 \rho = T^{00} =   \frac{m^2}{ 12 f^2} C_{ijk} C^{ijk} \to \frac{f^2}{2} (J^0)^2 \,,
\ee
where the arrow gives the result once specialized to the solution \pref{CijkForm}. Similarly
\be \label{BCTijsoln}
   T^{ij}  =   \frac{m^2}{ 2! f^2} \Bigl[   {C^i}_{\beta\lambda} C^{j\beta\lambda}  - \tfrac16 \gamma^{ij}   C_{\alpha\beta\lambda} C^{\alpha\beta\lambda} \Bigr]  \to \frac{f^2}{2}   \gamma^{ij}   (J^0)^2 \,,
\ee
where the final expression again specializes to \pref{CijkForm}. This nonzero stress energy is consistent with the energy density for the scalar side, which is given by  
\be\label{DualScalarTmn}
  T^{\mu\nu} =   f^2 (\partial^\mu \axion + J^\mu)(\partial^\nu \axion + J^\nu)     - \left[ \tfrac12 f^2 (\partial_\alpha \axion + J_\alpha)(\partial^\alpha \axion + J^\alpha)   + \frac{(m \axion - \cJ)^2}{2\cZ^2}  \right] g^{\mu\nu} \,.
\ee
Evaluating this at $m \axion = \cJ$ and $\partial_\mu \axion = 0$ therefore gives
\be\label{constaxTmn}
   T^{\mu\nu} =   f^2  J^\mu J^\nu     -  \tfrac12 f^2  J_\alpha J^\alpha   g^{\mu\nu} \,,
\ee
and so $T^{00}
= \tfrac12 f^2 (J^0)^2$ and $T^{ij} = \tfrac12 f^2 (J^0)^2 \gamma^{ij}$, in agreement with eqs.~\pref{BCT00soln} and \pref{BCTijsoln}.  

On the form side $C$ is pure gauge under the 3-form transformations and this is why $H = 0$. But it contributes a nontrivial value to $\widehat G$, and this generates a stress energy. On the scalar side this dualizes to a trivial solution where $\axion$ is a constant. There is nonetheless also a stress energy here due to the contact $J^2$ term in the lagrangian, showing how this contact term really is crucial in making both sides describe the same physics. 

The $J^2$ term represents a ubiquitous contact matter interaction on top of the usual $J^\mu \partial_\mu \axion$ coupling and is unusual from the perspective of axion phenomenology. One of its consequences is that the scalar energy density implied by \pref{DualScalarTmn}, which for a static spacetime is 
\be
  T_{00} =  \tfrac12 f^2 (\dot\axion + J_0)^2  + \tfrac12 f^2 (\nabla \axion + \bfJ)^2  + \frac{(m \axion - \cJ)^2}{2\cZ^2}  \,,
\ee
need {\it not} be minimized by the constant axion configuration when $J^\mu$ is nonzero. Indeed for $\bfJ = 0$ and $J^0 = \lambda n(t)$ with constant $\lambda$ and time-dependent charge density $n(t)$, in the absence of the scalar potential the energy is minimized by choosing $\dot \axion = - J_0 = \lambda n$, implying a linear growth of $\axion$ with time.

More generally, the axion minimizes its gradient energy by adjusting to ensure $\partial_\mu \axion + J_\mu = 0$ everywhere but this cannot be achieved if the matter current should have nonzero curl:\footnote{Notice the similarity between this and the discussion below eq.~\pref{divC}.} $\partial_\mu J_\nu \neq \partial_\nu J_\mu$. For instance, if the matter density is static but position dependent, $J^0 = \lambda n(\bfx)$ then minimizing $(\dot \axion + J_0)^2$ implies $\dot \axion(\bfx) = \lambda n(\bfx)$ is also position dependent and so the axion acquires a time-dependent spatial gradient even if $J^i = 0$. This makes it impossible to maintain zero stress energy in such cases. It is also inconsistent with an axion potential or with the requirement $m \axion = \cJ$ that minimizes the energy for static scalar current $\cJ$, implying an element of frustration in axion energetics.
 
The importance of this effect partially depends on the size of the current. In the examples obtained from higher dimension -- such as from the higher-dimensional theory described in \S\ref{sssec:CouplingStrengths} below, the coupling constant $\lambda$ is gravitational in strength: $\lambda \sim \cO(1/\Mp)$, suggesting that these current-current effects are likely mostly important in sources for which gravity provides a non-negligible part of the system's binding energy.

\subsubsection{Frustration}
\label{sssec:AxionMacro}

The field equation found from \pref{L2axionmassive2} by varying the axion is
\be \label{axioneq}
  \partial_\mu \Bigl[ \sqrt{-g} \, f^2 \left(\partial^\mu \axion + J^\mu \right) \Bigr] + \sqrt{-g} \left[ -\frac{m}{\cZ^{2}} \Bigl( m \axion - \cJ \Bigr)  + \frac{1}{4!} \epsilon^{\mu\nu\lambda\rho} \Omega_{\mu\nu\lambda\rho} \right] = 0 \,,
\ee
and so does not receive a contribution directly from the $J^\mu J_\mu$ term (which is, after all, axion independent). What {\it does} change however is the energy of a given solution and this can alter its stability and how the resulting field gravitates.

For instance let's assume that axion shifts are not anomalous (so the $\Omega$ term is not present) and that $\cJ$ is a constant (and so can be set to zero by shifting $\axion$ appropriately). For simplicity we also choose $f = \cZ = 1$, leaving the simplified equation
\be \label{axioneqq}
 (\Box - m^2 ) \axion + \nabla_\mu J^\mu   = 0 \,.
\ee
The axion is in principle not sourced by matter if the current to which it couples is covariantly conserved: $\nabla_\mu J^\mu = 0$. This result is unsurprising to the extent that the axion shift symmetry remains unbroken by the $J^\mu$ coupling. Conversely, if the current is not conserved then the wave solutions to $(\Box - m^2)\axion = 0$ can be driven in regions where $\nabla_\mu J^\mu \neq 0$.

To get a better intuition for the significance of the $J^\mu J_\mu$ interaction, consider a massless axion ($m=0$) coupled to a conserved current (such as baryon number) and suppose we seek the scalar configuration sourced by a static spherically symmetric matter distribution with $\bfJ = 0$ and $J^0 = \lambda n(r)$ for a coupling $\lambda$ and charge density $n(r)$. We assume the source weakly gravitates and so it suffices to work with a flat background metric. In this case \pref{axioneqq} implies $\nabla^2 \axion = 0$, which expresses in this instance conservation of the current $\nabla \axion + \bfJ$ that holds more generally in the $m \to 0$ limit. This implies in particular that any axion configuration that starts with vanishing $r^2 \partial_r \axion = 0$ at $r = 0$ integrates to give $R^2 \, \partial_r \axion = 0$ at the surface $r = R$ of the macroscopic source. The general solution to $\nabla^2 \axion = 0$ outside the source is $\axion(r) = \axion_0 + (Q_\ssA/r)$ where the source's axion `charge' is fixed by the boundary condition $Q_\ssA = - r^2 \partial_r \axion$ evaluated at $r = R$. Unsurprisingly we see in this way that the solution is simply a constant axion field and so the macroscopic source has $Q_\ssA = 0$. 

A key question asks whether this solution is stable, given that \pref{constaxTmn} implies constant axion fields carry nonzero energy density $\rho = \frac12 (J^0)^2 = \frac12 \lambda^2 n^2$ due to the contact interaction. In particular the possibility exists that the energy can be lowered for nonzero derivatives if $\partial_\mu\axion + J_\mu$ can be made small. With this in mind, suppose one zooms in to the source's centre in the vicinity of $r=0$ for a small enough region that the density $n$ can be regarded to be approximately constant. In this region the axion can locally lower its energy to $\rho = 0$ by acquiring a nonzero time-dependence with $\dot \axion = \lambda n$ and so $\axion = \axion_0 + \lambda n t$. This is indeed a solution to the field equation $\ddot \axion = 0$ that holds for homogeneous configurations when $m = 0$. 

Although suggestive, a solution with $\dot \axion = \lambda n$ is no longer zero energy if $n(r)$ depends on position because it implies a time-dependent spatial gradient $\partial_r \axion = \lambda t \partial_r n$, and this costs an ever-growing gradient energy because $\nabla \axion + \bfJ$ is nonzero. A similar energy cost to having a time-dependent axion also arises from the potential $\frac12 m^2 \axion^2$ when $m$ is nonzero. The minimal energy solution will not have zero energy and must be determined by balancing the mutual frustration introduced by seeking to minimize the sum of the kinetic energy $\frac12 (\dot \axion - \lambda n)^2$ and gradient energy $\frac12 (\nabla \axion)^2$ (and potential energy $\frac12 m^2 \axion^2$ if $m \neq 0$). 

If the gradient of $n$ is small enough relative to $m$ then it is the competition between kinetic and potential energy that dominates, and so the frustrated evolution can be modelled as being homgeneous, $\axion = \axion(t)$. In this case $\axion(t)$ must satisfy $\ddot \axion + m^2 \axion = 0$, which has as general solution:
\be
   \axion_c(t) =  \axion_0 \, \cos(m t) + \frac{v_0}{m} \, \sin(m t)  \,,
\ee
whose energy density
\be
 \rho = \tfrac12 (\dot \axion - \lambda n)^2 + \tfrac12 m^2 \axion^2 
 = \tfrac12(v_0^2 + m^2 \axion_0^2 + \lambda^2 n^2) - \lambda n \Bigl[  v_0 \, \cos(m t)   - m \axion_0 \, \sin(m t) \Bigr] 
\ee
satisfies
\be
 \dot \rho = \lambda m^2 n \, \axion_c \,.
\ee

The energy density is not constant for fixed $n$ because the axion-matter interaction causes the axion to exchange energy with the matter in the current. Energy conservation re-emerges once the stress energy and equations of motion for this matter are also included. Holding the current fixed assumes the matter energy is large enough not to be appreciably perturbed by the axion evolution. The energy averaged over a complete oscillation {\it is} conserved however, with 
\be
 \ol \rho = \frac{m}{2\pi} \int_0^{2\pi/m} \exd t \Bigl[ \tfrac12 (\dot \axion - \lambda n)^2 + \tfrac12 m^2 \axion^2 \Bigr] 
 = \tfrac12(v_0^2 + m^2 \axion_0^2 + \lambda^2 n^2) \,.
\ee
Because this is always greater than $\frac12 \lambda^2 n^2$ this shows that the energetics of competition between kinetic and potential energies does not favour time-dependence over constant configurations, despite the current-dependence of the kinetic energy.

A similar story with a different ending goes through if the gradient energy is not negligible. Acknowledging that time-dependence can be stimulated by nonzero $n$ suggests that the assumption the solution be static is too strong and we should instead seek solutions $\axion = \axion_c(r,t)$, that satisfy $(\Box - m^2) \axion_c = 0$. Writing $\axion = \alpha(r,t)/r$ this becomes
\be
   - \partial_t^2 \alpha +  \partial_r^2 \alpha   - m^2 \alpha = 0 
\ee
whose general solution $\alpha = \alpha_c(r,t)$ is a superposition of waves of the form $e^{-i\omega t + ik r}$ with $\omega = \pm \omega_k$ where $\omega_k := \sqrt{k^2+m^2}$. Writing
\be \label{GenAxionSoln}
   \axion_c(r,t) = \frac{1}{r} \int_{-\infty}^\infty \frac{\exd k}{\sqrt{4\pi \omega_k}} \Bigl[ \alpha_k \, e^{-i \omega_k t + ik r} + \alpha_k^* \, e^{i \omega_k t - i k r} \Bigr] 
\ee
the complex coefficients $\alpha_k$ are determined by the initial conditions $\alpha(r,t=0)$ and $\dot \alpha(r,t=0)$. 

The field energy becomes
\bea \label{geneng}
   E &=& 4\pi \int_0^\infty \exd r \, r^2 \rho = 2 \pi \int_0^\infty \exd r \left[ ( \dot \alpha_c - r\lambda n)^2 + m^2 \alpha_c^2 +   \left( \alpha_c' - \frac{\alpha_c}{r} \right)^2 \right]  \nn\\
   &=& E_0 + 2\pi \lambda^2 \int_0^R \exd r \, r^2 n^2 - 4 \pi \lambda \int_0^R \exd r \, r n(r) \, \dot \alpha_c(r,t) \,,
\eea
where over-dots denote $\partial_t$ and primes denote $\partial_r$ and 
\be
   E_0 = 2 \pi \int_0^\infty \exd r \left[ \dot \alpha_c^2 + m^2 \alpha_c^2 +   \left( \alpha_c' - \frac{\alpha_c}{r} \right)^2 \right] = \int_{-\infty}^\infty \exd k \, \alpha_k^* \alpha_k \, \omega_k \geq 0
\ee
is the energy when $\lambda \to 0$. One can ask whether there are solutions $\alpha_c(r,t)$ for which $\rho < \frac12 \lambda^2 n^2$, which \pref{geneng} shows requires $E_0$ to be smaller than the last term involving $n \, \dot \alpha_c$. The answer is clearly `yes' (provided $\alpha_c$ is time-dependent) if $\alpha_c$ is small enough because $E_0$ is quadratic in $\alpha_c$ while the last term is linear. Although the naive initial condition with $\alpha(r,0) = \dot \alpha(r,0) = 0$ corresponds to $\alpha_k = 0$ (and so $\alpha(r,t) = 0$ for all $t$), the solution is much different if (for instance) $\dot \alpha(r,0)$ is chosen to be nonzero and very small. 

In this last case the solution is time-dependent, with the oscillations near $r = 0$ described above for the homogeneous case here corresponding to $\alpha(0,t)$.  This oscillatory solution also stimulates wave emission -- and so also energy loss -- from the source, as is clearest in the $m \to 0$ limit, for which the general solution \pref{GenAxionSoln} can be written in the equivalent form $\axion(r,t) = r^{-1} [ f_1(r - t) + f_2(r + t) ]$ where $f_1(x)$ and $f_2(x)$ are otherwise arbitrary functions whose shape is determined by the initial condition $\axion(r,0)$ and $\dot \axion(r,0)$.

This strongly suggests that the presence of matter can spontaneously generate a nontrivial axion field despite the static solution with boundary condition $\lim_{r \to 0} \partial_r \axion = 0$ giving only the trivial solution. A full discussion of the endpoint of this evolution goes beyond the scope of this paper but we regard it to be of great interest to pursue.

\subsubsection{Cosmology}

Cosmology provides a second instance where the current-current contact interaction can modify axion evolution: in this case the gravitational response of the axion is important and so the possibility that currents can reduce the axion kinetic energy can change their contribution to the Friedmann equation and so alter the overall evolution, at least at early times. 

On the scalar side we solve \pref{axioneq} within a spatially flat FRW metric 
\be
   \exd s^2 = - \exd t^2 + a^2(t) \, \delta_{ij} \, \exd x^i \, \exd x^j \,, 
\ee
assuming that the current $J^\mu$ is conserved, $\partial_\mu (a^3 J^\mu) = 0$, when using the matter equations of motion (such as would for example be true for baryon number, post-recombination). Specialized to homogeneous and isotropic current distributions $J^0 = J^0(t) = \lambda n(t)$, $J^i = 0$ current conservation takes the familiar form $\partial_t (a^3 n) = 0$. If we again drop $\Omega$ and set $f = \cZ = 1$ and assume $\cJ$ is constant and so can be removed by shifting $\axion$, then the axion field equation \pref{axioneq} for homogeneous solutions $\axion(t)$ becomes
\be \label{axioneq3}
 \partial_t \Bigl[  a^3 \left(- \dot \axion + \lambda n \right) \Bigr] - a^3 m^2 \axion = 0 \quad \hbox{or} \quad  
\ddot \axion + 3 H \dot \axion + m^2 \axion = 0  \,,
\ee
where $H = \dot a/a$ is the usual Hubble parameter. Unlike in the static case current conservation is consistent with nonzero $\dot n$, but this in itself is no longer a driver of axion oscillations (unlike the flat-space situation described in Appendix \ref{App:DrivenOsc}). 

We see that when the mass term dominates the axion oscillations proceed in the same way they do in the absence of the axion-matter interaction. But the axion behaviour can differ from the naive treatment when the mass term is negligible. In this case the equation $\ddot \axion + 3H \dot \axion \simeq 0$ has two solutions: $\dot \axion = 0$ and $\dot \axion \propto a^{-3}$. In the usual telling of the axion story the $a^{-3}$ solution describes the process where Hubble friction causes the axion to freeze at a constant value. During this freezing process its gravitational effects on the evolution of the universe are felt through the axion energy's contribution to the Friedmann equation, which adds a rapidly falling $\rho_{\rm ax} \propto \dot \axion^2 \propto a^{-6}$ contribution to $H^2$. In the presence of the current-current interaction, however, the axion energy is instead proportional to $(\dot \axion - \lambda n)^2$ and so having $\dot \axion = \lambda n$ is both consistent with the field equations (since both $\dot \axion$ and $n$ are proportional to $a^{-3}$) and completely removes the axion from the Friedmann equation. This continues until the axion mass eventually becomes non-negligible and the oscillation energy contributes. Because the energy difference falls like $a^{-6}$ it likely only plays a role at very early times. A more detailed study of the cosmology of this model, including fluctuations, is underway. 

\section{Naturalness in the eye of the dual}
\label{ssec:NaturalnessDual}

One of the points of the previous section is that duality maps the scalar potential onto a function of the kinetic energy of the dual field. This necessarily changes the number of derivatives that appear in the interaction and this has important implications when it is the derivative expansion that justifies the use of semiclassical methods (as is essentially always the case for gravitational interactions \cite{Burgess:2003jk, Burgess:2020tbq}). This section explores how this observation affects standard power-counting arguments and the kinds of interactions to be expected at low energies (elaborating an argument made in \cite{Burgess:2023ifd}). 

\subsection{Power counting review}
\label{sssec:PowerCountingReview}
 
The role of the derivative expansion for controlling the semiclassical limit in cosmology is a general consequence of the modern understanding of how to compute reliably with nonrenormalizable theories \cite{Weinberg:1978kz} and we here provide a minor extension of the arguments of ref.~\cite{Burgess:2009ea, Adshead:2017srh} to include bosonic form fields. 

Consider therefore a collection of $N_s$ scalar fields, $\theta^i$, and $N_\ssF$ form fields $B_{\mu_1\cdots \mu_p}$ coupled to other fields (including gravity) with gravitational strength. Here both scalar and form fields have their dimension scaled out to make them dimensionless and the assumption of gravitational strength corresponds to the choice of using the Planck scale to normalize their kinetic terms. Writing the field strength $G = \exd B$, the effective lagrangian for these fields written as a derivative expansion has the schematic form
\ba \label{Leffdef}
 - \frac{ \cL_\eff}{\sqrt{-g}} &=& v^4 V(\theta) + \frac{M_p^2}{2}
 \, \Bigl[  W(\theta) \, R
 + \cG_{ij}(\theta) \, \partial_\mu \theta^i
 \partial^\mu \theta^j + H_{ab}(\theta) G^a_{\mu_1\cdots\mu_p} G^{b\mu_1\cdots \mu_p}  \Bigr] \nn\\
 && \quad +\Bigl[ A(\theta) (\partial \theta)^4 + B(\theta)
 \, R^2 + C(\theta) \, R \, (\partial \theta)^2
 + D(\theta) \, G^4  + \cdots \Bigr] \\
 && \qquad\qquad\qquad\qquad + \frac{E(\theta)}{M^2} \, (\partial \theta)^6
 + \frac{F(\theta)}{M^2} \, R^3 + \frac{K(\theta)}{M^2} \, G^6 + \cdots \,,\nn
\ea
where terms with up to two derivatives written explicitly and the rest written more schematically, inasmuch as the ellipses in the second line contains all possible independent curvatures, scalars and form field-strengths involving six derivatives, and so on on the last line for even higher derivatives. The explicit mass scales $v$, $M_p$ and $M$ are extracted so that the functions $V(\theta)$, $W(\theta)$, $A(\theta)$, $B(\theta)$ {\it etc}, are dimensionless. Notice that six-derivative terms are generically suppressed by a scale $M \ll M_p$ \cite{Burgess:2003jk} because it is the lowest mass that wins in a denominator when contributions of multiple virtual states are summed. For cosmological applications one usually has in mind $V \simeq v^4 \ll M^4$ when $\theta \simeq \cO(1)$, where $v^2/M_p \sim H$ sets the scale of the metric's background curvatures. 

Again expanding about a classical solution,
\be \label{semiclassexp}
 \theta^i(x) = \vartheta^i(x) + \frac{\phi^i(x)}{M_p} \,, \quad 
 B^a_{\mu_1 \cdots \mu_p}(x) = \cB^a_{\mu_1 \cdots \mu_p}(x) + \frac{b^a_{\mu_1 \cdots \mu_p}}{M_p} \,,
 \quad  
 g_{\mu\nu} (x) = \hat g_{\mu\nu} (x) +
 \frac{h_{\mu\nu}(x)}{M_p} \,,
\ee
allows the lagrangian in eq.~\pref{Leffdef} to be written as 
\be \label{Leffphih}
 \cL_\eff = \hat \cL_\eff + M^2 M_p^2 \sum_{n}
 \frac{c_{n}}{M^{d_{n}}} \; \cO_{n} \left(
 \frac{\phi}{M_p} , \frac{ b_{\mu_1\cdots \mu_p}}{M_p} , \frac{ h_{\mu\nu}}{M_p} \right)
\ee
where $\hat\cL_\eff = \cL_\eff(\vartheta,\cB_{\mu_1 \cdots \mu_p}, \hat g_{\mu\nu})$ and the interactions, $\cO_{n}$, involve $N_n = N^{(\phi)}_n  + N^{(\ssB)}_n + N^{(h)}_n \ge 2$ powers of the fields $\phi^i$, $b^a_{\mu_1 \cdots \mu_p}$ and $h_{\mu\nu}$. The parameter $d_{n}$ counts the number of derivatives appearing in $\cO_n$, the coefficients $c_n$ are dimensionless and the prefactor $M^2 M_p^2$ ensures the kinetic terms (and so also the propagators) are $M$ and $M_p$ independent. 

Following \cite{Burgess:2009ea, Adshead:2017srh}, we assign $M$, $M_p$ and $v$ dependence to the coefficients $c_n$ (for $d_n \ne 2$) so that eq.~\pref{Leffphih} captures the same dependence as is obtained when the lagrangian \pref{Leffdef} is expanded using \pref{semiclassexp}. In the simplest case derivatives of the background and fluctuations have the same size (such as for fluctuations in cosmology with modes at horizon exit, where $k/a \sim H$). We relax this assumption in the next section. For $d_n > 2$ this implies $c_n$ is given by 
\be \label{cndgt2h}
 c_n = \left( \frac{ M^2}{ M_p^2} \right) g_n
 \qquad \hbox{(if $d_n > 2$)} \,,
\ee
where $g_n$ is order unity, and for terms with no derivatives --- {\it i.e.}~those coming from the scalar potential, $V(\theta)$ --- we have
\be \label{cndeq0}
 c_n = \left( \frac{v^4}{M^2 M_p^2} \right) \lambda_n
 \qquad \hbox{(if $d_n = 0$)} \,,
\ee
where the dimensionless couplings $\lambda_n$ are independent of $M_p$ and $M$. These choices can be relaxed but have the virtue that they allow the power-counting argument made below to be based purely on dimensional analysis. 

We remark in passing that this last parameterization is equivalent to writing the scalar potential in the form
\be \label{Vvslambda}
 V(\phi) = v^4 \left[ \lambda_0 + \lambda_2 \left(
 \frac{\phi}{M_p} \right)^2 + \lambda_4 \left(
 \frac{\phi}{M_p} \right)^4 + \cdots \right] \,,
\ee
and so $V$ ranges through values of order $v^4$ as $\phi^i$ range through values of order $M_p$. In particular this makes the natural scale for the scalar masses of order $m \simeq \sqrt{V''}/M_p \simeq v^2/M_p \simeq H$, where we assume the natural value for the Hubble scale: $H \simeq \sqrt{V}/M_p \simeq v^2/M_p$. This makes explicit an old observation \cite{Albrecht:2001xt}: having light scalars need not introduce new naturalness problems {\it beyond the question of why $v \ll M, M_p$ in the first place} ({\it i.e.}~the cosmological constant problem), provided those scalars are gravitationally coupled. 

Our goal is to identify how observables depend on both the small ratios $H/M$ and $H/M_p$ when evaluated perturbatively using the above semiclassical expansion. To this end we expand $\cL_\eff = \Bigl( \hat \cL_\eff + \cL_0 \Bigr) + \cL_{\rm int}$, where $\cL_0$ contains terms quadratic in the fluctuation fields $\phi^a$, $b^a_{\mu_1 \cdots \mu_p}$ and $h_{\mu\nu}$and lump all cubic and higher terms in to $\cL_{\rm int}$. We evaluate the path integral perturbatively in $\cL_{\rm int}$ in the usual way, and focus our attention on the size of a Feynman graph having a total of $\cE$ external lines, with external lines and derivatives of background fields characterized by the single low-energy scale $H$. We track the powers of $M$, $v$ and $M_p$ coming from the vertices in the Feynman rules and determine the $H$-dependence on dimensional grounds (see \cite{Burgess:2009ea, Adshead:2017srh} for details).\footnote{Since regularizing UV divergences with cutoffs makes dimensional arguments more difficult we quote the result that is obtained assuming dimensional regularization is used as the UV regulator.}

For a graph with $L$ loops and $V_n$ vertices of type `$n$', involving $d_n$ derivatives whose external legs are amputated\footnote{For cosmology correlation functions are of more practical interest, for which external lines are not amputated. Denoting the corresponding result by $\cC_\cE$ these scale relative to the amputated amplitude $\cA_\cE$ like $\cC_\cE \simeq \cA_\cE H^{2\cE-4}$.} the above prescription leads to the following dependence on scales:
\bea \label{PCresult}
 \cA_\cE (H) &\simeq& H^2 M_p^2 \left( \frac{1}{M_p} \right)^\cE
 \left( \frac{H}{4 \pi \, M_p}
 \right)^{2L} \left[ \prod_{d_n = 2}  c_n^{V_n} \right]  \nn\\
 && \qquad\qquad \times \prod_{d_n = 0} \left[ \lambda_n \left( \frac{v^4}{H^2 M_p^2}
 \right) \right]^{V_n}
 \prod_{d_n \ge 4} \left[ g_n \left( \frac{H}{M_p}
 \right)^2 \left( \frac{H}{M}
 \right)^{d_n-4} \right]^{V_n}  \,.
\eea
This expression gives the size of any particular Feynman graph and provides the basis for understanding the size of the implications (and so also the relative importance) of different terms in $\cL_{\rm eff}$. Several important conclusions can be drawn immediately:
\begin{itemize}
\item The contribution of all interactions involving $d_n > 2$ derivatives are always suppressed by powers of $H/M$ and $H/M_p$ and so it is the small size of these ratios ({\it i.e.}~the derivative expansion) that allows these contributions to be small (and so be neglected at leading order).
\item In the small-$H$ regime loops are systematically suppressed by powers of $H/M_p$ and this is why the classical approximation becomes a good approximation for gravity at low energies. In particular, using 2-derivative interactions at higher loop order competes with the use of higher-derivative interactions at lower loop order (and this is why it is possible for higher-derivative interactions to renormalize UV divergences coming from loops involving {\it e.g.}~just the 2-derivative interactions of \pref{Leffdef}). As a corollary, should $H/M_p$ ever be large enough to make higher derivatives competitive with 2-derivative interactions within a classical approximation then higher loops are also generically important and the validity of semiclassical methods themselves is generically suspect.
\item The contributions of interactions involving no derivatives ($d_n = 0$) are {\it dangerous} in the small-$H$ limit because they involve $H$ in the denominator rather than the numerator. The scalar potential can systematically undermine the validity of the classical approximation at low energies and the famous naturalness problems of scalar potentials are a special case of this observation. It is for this reason that shift symmetries are often invoked to suppress the scalar potential.\footnote{An important situation where scalar interactions do not threaten semiclassical methods is when $H \simeq v^2/M_p$ (as is usually the case in cosmological applications) because when this is true the enhancement factor $v^4/(H^2 M_p^2)$ is order unity. For such models the main question becomes why $v$ is small.}
\item The contribution of 2-derivative interactions ($d_n = 2$) are neither enhanced or suppressed at low energies. This is why the nonlinearity of General Relativity cannot be neglected in the low-energy limit. It also shows that it is the two-derivative interactions of the other fields -- the first line (excluding the scalar potential) of \pref{Leffdef} -- that want to compete with GR at low energies (making these interactions perhaps the most interesting places to search for deviations from GR at low energies, as has been argued for multiple-scalar models in \cite{Brax:2023qyp, Smith:2024ayu, Smith:2024ibv, Smith:2025grk}).
\item Also unsuppressed are contributions involving high powers of the fields (as opposed to their derivatives), given that large fields do not in themselves give large energies with the form assumed in \pref{Leffdef} for the scalar potential. The low-energy limit need not be inconsistent with large -- even trans-Planckian -- fields, as the self-consistency of supersymmetric flat directions illustrate.  
\end{itemize}

It is the observation that the derivative expansion is intrinsically related to the semiclassical expansion in gravity that has interesting implications for dual theories because duality can exchange higher-derivative interactions for zero-derivative interactions in the scalar potential, such as happens for scalar masses in \S\ref{sssec:MassiveCase} (and more generally -- see below). 
 
\subsection{Power counting \& duality for axions}
\label{ssec:SPot}

The previous section shows how power-counting arguments (as always) organize interactions according to their number of derivatives because derivative expansions are ultimately reponsible for the validity of the semiclassical approximation (for gravity at least). This is equally true on either side of the duality to the extent that both sides involve scalars, gauge forms and the metric in the way assumed by the power-counting arguments of \S\ref{sssec:PowerCountingReview}. This seems to be borne out by duality relations like \pref{GSaddle} involving the same number of derivatives on both sides: sensibly enough the low-energy expansion is equally valid on both sides of the duality relation.  

Slightly more worrisome are duality relations like \pref{GSaddlem} that mix terms involving different numbers of derivatives. It is the presence of such terms that allows duality to map 2-derivative terms like $H_{\mu\nu\lambda\rho}H^{\mu\nu\lambda\rho}$ to zero-derivative terms like $m^2 \axion^2$. Changing the number of derivatives in the interaction mixes up where these terms appear in the power-counting arguments. Zero-derivative interactions are generically dangerous since they {\it undermine} the low-energy expansion, so the only way that they can be equivalent to the effects of derivative terms is if their effective couplings -- the $\lambda_n$'s of eq.~\pref{cndeq0} or \pref{Vvslambda} -- are anomalously small. In particular the mass term and the kinetic term are equally suppressed at low energies for small masses if we regard $m$ as effectively counting as a derivative when power-counting.

Although perhaps not surprising for the quadratic mass term this observation has more implications for more complicated interactions. For instance suppose we add $\delta \cL = W(X)$ of the form
\be \label{Wquart}
   W = -\cJ X + \frac{\cZ^2}2 \, X^2 + \frac{2c_3}{3M^2} \, X^3 + \frac{c_4}{4M^4} \, X^4 + \cdots
\ee
to the lagrangian where $X := \frac{1}{4!} \epsilon^{\mu\nu\lambda\rho} H_{\mu\nu\lambda\rho}$ (and so $X^2 = - \frac{1}{4!} H_{\mu\nu\lambda\rho}H^{\mu\nu\lambda\rho}$). The factors of the UV scale $M$ are chosen so the coefficients $c_i$ are dimensionless when $H_{\mu\nu\lambda\rho}$ is canonically normalized. Eq.~\pref{Wquart} reproduces our earlier discussion of massive scalars -- {\it c.f.}~eq.~\pref{L1hatG} -- if only the first two terms are kept. 

For non-quadratic $W$ the integral over $H$ is not gaussian, but within the classical approximation the saddle point \pref{HSaddle} is modified to
\be \label{alphavsWX}
    \left( \frac{\partial W}{\partial X} \right)_{H = \cH} = -m\, \axion \,,
\ee
which agrees with \pref{HSaddle} when $W = -\cJ X + \frac12\, \cZ^2\, X^2$. For the choice \pref{Wquart} this instead becomes
\be
  -\cJ + X\left( \cZ^2 + \frac{2c_3}{M^2} X + \frac{c_4}{M^4} X^2 + \cdots \right) \simeq -m \, \axion
\ee
and so
\be
   X \simeq \frac{\cJ - m\, \axion }{\cZ^2}  - \frac{2c_3}{M^2\cZ^6}\left( \cJ - m \axion \right)^2 
    + \cdots \,.
\ee

Once used in the lagrangian this shows how non-quadratic pieces of $W$ map over to non-quadratic contributions to the scalar potential for $\axion$ in the dual lagrangian $\cL_2$.  In particular the axion potential becomes
\be
  V(\axion) = - W(X) - m \axion X = \frac1{2\cZ^2} \left(\cJ - m\, \axion \right)^2 - \frac{2c_3}{3M^2\cZ^6}  \left(\cJ - m\, \axion \right)^3 + \cdots \,. 
\ee
Notice this potential shares the usual Legendre property
\be \label{dVdalphavsX}
  \frac{\partial V}{\partial \axion} = -m X - \left(   \frac{\partial W}{\partial X} + m \axion \right) \frac{\partial X}{\partial \axion} = -mX\,,
\ee
where the last equality uses \pref{alphavsWX}, so even if non-quadratic terms introduce new stationary points for $V(\axion)$ (or shifts the positions of old ones) eq.~\pref{dVdalphavsX} ensures $X = 0$ for {\it all} of them. 

Notice also that the potential depends on $m$ and $\axion$ systematically only through the combination $m \axion$ and as a result a term proportional to $\axion^n$ comes suppressed by a power of $(m/M)^n$ relative to what would naively be expected on dimensional grounds for $V(\axion)$.  This is how the dual reproduces the same low-energy physics despite the power-counting of a potential and kinetic terms being superficially very different. This means that scalar potentials for dual systems can be very small compared with generic power-counting arguments for scalars. In particular, in the limit $M \to \infty$ in which it suffices to drop anything inversely proportional to $M$ the potential is simply quadratic in $\axion$ (as opposed to being a trigonometric function of $\mfa$).\footnote{Having a quadratic potential can be consistent with the axion target space being periodic if the potential arises as the lowest branch of a sum of quadratic potentials centered about each distinct vacuum. Such a construction introduces cusps where different branches meet but the fields required to probe these cusps satisfy $m \axion \sim M^2$ and so are only accessible when the low-energy limit is breaking down.} 

How can this be consistent? After all, general power-counting arguments are usually what drive our expectation for what the generic size of loop-generated effects in the scalar potential should be in the first place. The point is that the size usually expected for corrections to the lagrangian assumes generic couplings, such as if the dimensionless couplings $\lambda_n$ or $g_n$ are all order unity once all important scales are made explicit. But these dimensionless couplings cannot all be order unity on both sides of the duality transformation. If they are order unity on the form-field side then they are parameterically suppressed by powers of $m/M$ on the scalar side, in just the right way to ensure that the factors of $\lambda_n$ in the $d_n=0$ term of \pref{PCresult} capture the same behaviour as for the dual (given that $m \sim H$). \emph{Our prejudices about what size couplings are generically natural need not commute with duality transformations.} 

We emphasize that these arguments do not mean that it is impossible to have a complicated potential for a scalar that has a dual formulation. It is just that the complicated function of the field strength this requires on the form-field side is then equally complicated, and so when the shape of the potential becomes important the predictions made on the dual side cannot rely on the usual derivative expansion to control the semiclassical limit. The QCD axion is likely an example of this type, for which both the dual 4-form field strength and the scale $M$ are of order $\Lambda_\QCD$ (see for example the discussions in \cite{Burgess:2023ifd, Dvali:2005an}). It might be that the existence of a dual scalar formulation can provide a way to {\it define} what we mean by semiclassical physics for form field-strengths that are not small compared to the UV scale $M$.    

\subsubsection{Matter-coupling priors}

Another place where duality can change priors that are driven by naturalness arguments is the classification of the lowest-dimension interactions that a scalar axion has with matter fields. These can be very different depending on whether the axion is formulated as a generic scalar Goldstone boson or the dual of a form field. Power-counting tells us that the most important couplings at low energy have the lowest mass dimension, but the lowest-dimension interactions with matter take different forms on either side of the duality relation. 

For instance a canonically normalized scalar axion can have effective couplings to Standard Model particles of the schematic form 
\bea
   &&\hbox{dim 3}: \qquad\qquad \axion \,\cO_2  \nn\\
   &&\hbox{dim 4}: \qquad\qquad \axion^2 \cO_2 \,, \quad \axion \cO_3 \\
   &&\hbox{dim 5}: \qquad\qquad \axion^3 \cO_2 \,, \quad \axion^2 \cO_3 \,, \quad \axion \cO_4 \,, \quad \partial_\mu \axion J^\mu_3  \nn
\eea
{\it etc} where $\cO_n$ is a Lorentz scalar, gauge-invariant operator with dimension (mass)${}^n$ and $J^\mu_n$ is a Lorentz vector, gauge invariant operator with dimension (mass)${}^n$ and so on. The lagrangian density is obtained from these by multiplying by a factor of $\sqrt{-g}$. 

For the Standard Model $\cO_2 = H^* H$, there are no gauge invariant $\cO_3$ operators, while $\cO_4$ contains all of the kinetic terms for all SM fields plus the Yukawa couplings and terms like $\epsilon^{\mu\nu\lambda\rho}\Tr[ F_{\mu\nu}F_{\lambda\rho}]$. Similarly $J^\mu_3$ involves fermion bilinear operators $\ol\psi \, \mfm^\mu \psi$, for some Dirac and flavour matrix $\mfm^\mu$, and Higgs operators like $H^\dagger D^\mu H$.  

By contrast, the most general couplings of the canonically normalized fields $C_{\mu\nu\lambda}$ and $B_{\mu\nu}$ to the Standard Model that are gauge invariant under the form symmetries and Standard Model gauge symmetries out to dimension five are those captured in the cross terms involving the Chern Simons coupling within $\widehat G$ -- as in \pref{GChern} -- and
\bea
   &&\hbox{dim 4}: \qquad\qquad  \epsilon_{\mu\nu\lambda\rho} H^{\mu\nu\lambda\rho} \cO_2 \nn\\
   &&\hbox{dim 5}: \qquad\qquad \epsilon_{\mu\nu\lambda\rho} \widehat G^{\mu\nu\lambda} J^\rho_3 \,. 
\eea 
Notice none of these involve the spacetime metric (apart from any metrics contained within the definitions of $\cO_2$ and $J^\mu_3$) once made generally covariant. The dimension-4 interaction dualizes to $m \axion \, \cO_2$ in which the factor of $m$ raises the effective dimension by one (and so suppresses it by a factor of the small axion mass). Eq.~\pref{GSaddlem} shows that the dimension-5 interaction dualizes to the interaction $f^2 \partial_\mu \axion \, J^\mu_3$, though also including a new $J^\mu J_\mu$ term. Notice also that $f$ is in the numerator in this dual, not the denominator.

In unitary gauge (where we gauge fix $B=0$) the only part of the dimension-5 interaction that survives is the coupling between $C$ and $J_3$. For matter at rest, for which $J^\mu_3 \propto \delta^\mu_0 \mfn$, where $\mfn$ is a particle number density, this means that static matter distributions act as sources for the spatial components $C_{ijk}$. 

To summarize, although classical solutions map into one another under duality, the novelty of using a dual form description rather than a scalar description of an axion is that power-counting changes our naturalness prejudices for the size of interactions in the scalar potentials and for the kinds of possible axion-matter couplings. 

\subsection{$F(R)$ gravity: another derivative-nonderivative interchange}

There are other examples of classical equivalence wherein derivative and nonderivative terms are interchanged, a famous one being the equivalence between $F(R)$ gravity and a scalar-tensor field theory. $F(R)$ gravity is a theory whose action involves a more complicated function of the Ricci scalar, $R$, than does the Einstein-Hilbert action. Although the issues that arise in this case are similar in some ways to the discussion of \S\ref{ssec:NaturalnessDual}, it is also less clean because power-counting within the effective description on the gravity side is usually not fully specified. 

\subsubsection{$F(R)$ metric-scalar duality}

For example consider the path integral
\be \label{FRxi0}
   \Xi[\psi] = \int \cD g_{\mu\nu} \; e^{i S_1[g, \psi]} \,,
\ee
with action
\be \label{FRS1}
   S_1 = - \int \exd^4x \; \sqrt{-g} \Bigl\{ F(R) + L_m(\psi, g_{\mu\nu}) \Bigr\}
\ee
where $L_m$ is the action for some matter field $\psi$, that can either be a propagating field or an external current used to generate correlators of the metric. The function $F(R)$ is to be specified and much of what follows can be done for relatively general choices for $F(R)$. When push comes to shove we choose the simplest nontrivial example, 
\be
   F(R) = \tfrac12 M_p^2 R - \tfrac12 \xi R^2 \,, 
\ee
for couplings $\xi$ and $M_p$. This choice arises in practice in the Starobinsky model of inflation \cite{StarobinskyInflation}, whose phenomenological success requires ($i$) trusting semiclassical methods for solutions for which $M_p^2 R$ is similar in size to $\xi R^2$, and $(ii)$ choosing the parameter $\xi$ to satisfy $\xi \sim 10^8$. 

A dual description in the case of quadratic $F(R)$ can by found by uncompleting the square to write
\be
   \Xi[\psi] = \int \cD g_{\mu\nu} \cD \sigma\; e^{i S_0[g , \psi , \sigma]} \,,
\ee
with action\footnote{We use Weinberg's curvature conventions, which differ from MTW conventions only by an overall sign in the definition of the Riemann tensor.}
\be \label{FRS0action}
   S_0 = - \int \exd^4x \; \sqrt{-g} \left\{\tfrac12 M_p^2 R + \frac{ \sigma^2}{2\xi} +  \sigma R + L_m(\psi, g_{\mu\nu}) \right\} \,.
\ee
This is equivalent to the previous formulation because the gaussian integral over $\sigma$ can be performed explictly, with saddle point $\sigma_s = -\xi R$, leading back to \pref{FRxi0} and \pref{FRS1}. The dual instead rescales the metric in the action \pref{FRS0action} to put it into Einstein frame, defining
\be
   \hat g_{\mu \nu} := e^{2A} g_{\mu\nu} = f(\sigma) \, g_{\mu\nu} \qquad \hbox{with} \qquad f := 1 + \frac{2\sigma}{M_p^2} \,.
\ee
Using the 4D identity
\be \label{expnAR}
  \sqrt{-\hat g} \; \hat g^{\mu\nu} \hat R_{\mu\nu} 
  = e^{2A} \sqrt{-g} \, g^{\mu\nu} \Bigl[ R_{\mu\nu} - 6 \,\partial_\mu A \, \partial_\nu A \Bigr] + 3 \Box \Bigl[ e^{2A} \Bigr] \,,
\ee
the action \pref{FRS0action} becomes (dropping a surface term)
\begin{ignore}
\be \label{FRS0action2}
   S_0 = - \int \exd^4x \; \sqrt{-\hat g} \left\{\tfrac12 M_p^2\, \hat g^{\mu\nu} \left[ \hat R_{\mu\nu}   + \frac32 \left( \frac{\partial_\mu f \, \partial_\nu f}{f^2} \right) \right]- V(\sigma) + L_m(\psi, \hat g_{\mu\nu}/f) \right\} \,.
\ee
\end{ignore}
\be \label{FRS0action2c}
   S_0 = - \int \exd^4x \; \sqrt{-\hat g} \left\{\tfrac12 M_p^2\, \hat g^{\mu\nu} \Bigl[ \hat R_{\mu\nu}   +   \partial_\mu \phi \, \partial_\nu \phi \Bigr]- V(\phi) + L_m(\psi, \hat g_{\mu\nu}/f) \right\} \,,
\ee
where
\be
    \phi := \sqrt{\frac32} \, \ln \left( 1 + \frac{2\sigma}{M_p^2} \right) \quad \hbox{and so} \quad
    \sigma = \tfrac12  M_p^2 \Bigl( e^{\sqrt{\frac23} \phi } -1 \Bigr) \,.
\ee
The potential appearing in \pref{FRS0action2c} is
\be
    V 
    = \frac{1}{2\xi} \, \frac{\sigma^2}{(1 + 2 \sigma/M_p^2)^2} 
    = \frac{M_p^4}{8\xi} \Bigl( 1 -  e^{-\sqrt{\frac23} \phi }\Bigr)^2  \,.
\ee
This potential is bounded from below when $\xi$ is positive and satisfies $V \ll M_p^4$ (a necessary condition for trusting EFT methods) in the regime when $\xi \gg 1$.

\subsubsection{EFT issues}

On the scalar side this theory fits into the class of EFTs for which the power-counting of \S\ref{sssec:PowerCountingReview} applies, with the scale governing the potential in this EFT picture given by 
\be
  v \sim \frac{M_p}{\xi^{1/4}} \qquad \hbox{and so} \qquad
  H \sim \frac{v^2}{M_p} \sim \frac{M_p}{{\sqrt\xi}\;} \,,
\ee
so the generic loop-counting parameter appearing in \pref{PCresult} is
\be
   \left( \frac{H}{4\pi M_p} \right)^2 \sim \frac{1}{16\pi^2 \xi} \,.
\ee

The model is mute about the size of the scale $M$ describing the scale of the next level of physics (perhaps the string scale) that was integrated out to obtain the lagrangian \pref{FRS1} in the first place, but this can live only in a specific $\xi$-dependent range
\be
   \frac{M_p}{{\sqrt\xi}\;} \ll M \ll M_p \,. 
\ee
On one hand the right-most of these inequalities follows because the theory does not in itself UV complete gravity. On the other hand the left-most inequality follows because in cosmology $H$ must be a low-energy scale in order to trust the adiabatic approximation that underlies the use of EFTs in cosmology \cite{Burgess:2003jk, Burgess:2020tbq}. It is also required by the condition that a higher-order term $F(R) \ni R^3/M^2$ not compete with the $\xi R^2$ term. 

When expanding about $\phi = 0$ the dimensionless couplings $\lambda_n$ from eq.~\pref{cndeq0} or \pref{Vvslambda} are order unity (as was assumed in the explicit estimates for this model of \cite{Burgess:2009ea}) but when the classical background field satisfies $\bar\phi \gg 1$ (and so $\sigma \gg M_p^2$) the potential becomes very flat and so its derivatives are small in this regime because the magic of exponential potentials makes $\lambda_n \sim e^{- \sqrt{\frac23} \bar\phi} \ll 1$ for all $n$. For inflationary applications it is this suppression that drives the slow roll approximation, with for example $\epsilon = - \dot H/H^2 = \frac12 (V'/V)^2 \simeq \frac13 e^{-\sqrt{\frac83} \bar \phi}$ when $\bar \phi \gg 1$, showing that slow roll in this case is driven by the regime in field space (as opposed to making a tuned choice for parameters in a potential  \cite{Burgess:2001vr}).    

The metric side of the theory -- at least as used in inflation -- does not fit as nicely into the power-counting framework described in  \S\ref{sssec:PowerCountingReview} and for this reason when push comes to shove it is the scalar side of the duality that is actually used when assessing the size of theoretical errors. One problem is the parameter choices used are inconsistent with thinking of $F(R)$ as being perturbatively close to $M_p^2 R$ because inflation requires $\xi H^2 \sim M_p^2$. The inflationary solution requires balancing the $R$ term against the $R^2$ term and the inability to treat $R^2$ perturbatively is also why the curvature-squared term cannot be discarded as `redundant' (as is often done for EFT interactions involving the Ricci scalar \cite{Burgess:2003jk, Burgess:2020tbq}). 

The issue dealt with by power counting is how to handle the higher-dimension interactions that are {\it not} written explicitly in \pref{FRS1} and in particular self-consistently justify their neglect even in the presence of quantum fluctuations. This is usually controlled by the derivative expansion -- as sketched in \S\ref{sssec:PowerCountingReview} -- but this must be abandoned when allowing $R^2$ to compete with $R$. One might press on in any case and imagine expanding $F(R)$ about its classical saddle-point solution $R = R_c$ and then quantifying the deviations in terms of a controlled expansion -- along the lines that can be done \cite{Babic:2019ify} for the DBI action (which also goes beyond straight-up derivative expansions) -- but (to our knowledge) it is not yet known how to do so for \pref{FRS1}.  
 
\section{When the dual field does not propagate}
\label{ssec:NonPropForms}

4-form field strengths like $H_{\mu\nu\lambda\rho}$ are an example of a field that does not propagate and a natural point of view to take about these fields is: why include such fields at all?

The main reason for doing so is because they bring to the low-energy effective theory the news about how the UV completion responds to spacetime topology. How 4-forms appear in a low-energy theory also has implications for the structure of the low-energy scalar potential and consequently on naturalness issues, and this is implicit in their frequent invocation in discussions of the cosmological constant problem (see \cite{Kaloper:2025goq} for a recent attempt). 

In this section we briefly survey several examples and describe how some of these can also potentially impact low-energy axion behaviour (particularly within extra-dimensional models, since these often both involve 4-forms and axions arising in their dual formulation in terms of forms). We briefly describe the appearance of 4-forms in several approaches to the cosmological constant problem and argue in particular that relaxation models provide yet another way in which generic scalars can differ from those that arise as duals: for generic scalars the same relaxation that suppresses the Dark Energy density also suppresses the axion potential while the same is not true of axions that arise as the duals of 2-form potentials. 

\subsection{4-forms and extra-dimensional topology}
\label{ssec:NonPropFormsTopo}

The discussion of \S\ref{sssec:MassiveCase} shows how a 4-form not describing a propagating particle is crucial when thinking of a scalar field acquiring its mass through a Higgs mechanism within the dual formulation. Keeping such non-propagating fields can nevertheless be the difference between success and failure when describing a low-energy EFT because they are often the messengers that bring to the low-energy world the news that the underlying physics can be sensitive to topology. 

The physics of two spatial dimensions provide a number of concrete examples of this type, including some with experimental implications (like Quantum Hall systems).  A gauge field, $A_m$, in 2+1 dimensions governed by a Chern-Simons lagrangian, $\epsilon^{mnp} A_m F_{np}$ provides a simple example of this type. In this case the Maxwell lagrangian $F_{mn} F^{mn}$ is higher-derivative and so the field does not propagate a particle state if the Maxwell field is treated perturbatively relative to the Chern-Simons term. It can nonetheless have important physical effects if particles couple to the gauge field since the resulting flux lines can alter the particle statistics, leading to anyons. Such particles can describe the low-energy quasiparticle excitations about Quantum Hall vacua and the associated statistics gauge fields can capture the sensitivity of Quantum Hall systems to sample topology \cite{QHReviews}. 

The remainder of this section describes a similar example involving 4-forms in 3+1 dimensions that can have more direct implications for particle physics and cosmology. This example turns out to be a simple representative of a common phenomenon in string vacua, wherein 4D 4-form fields very often play an important role in bringing the news to the low-energy theory about extra-dimensional flux topology and the associated breaking or preserving of supersymmetry. As argued in \cite{Bielleman:2015ina} for flux-stabilized string compactifications leading to an effective supersymmetry in the 4D theory the contributions of 4-forms are described by the supergravity's auxiliary fields. These 4-forms are duals of higher-dimensional fluxes some of whose components are wrapped about cycles in the extra dimensions. Having 4-forms in 4D be related to the auxiliary fields of supersymmetry proves to be the rule and not the exception. 

\subsubsection{Salam-Sezgin 6D supergravity} 
\label{sssec:SalamSezgin}

Perhaps the simplest example of how low-energy 4-forms can play an important role in 4D comes when the 4D theory arises as the low-energy limit of a compactifications of 6D supergravity \cite{Burgess:2015lda}. The theory describes 4D branes situated at points within the two extra dimensions of a compactified 6D theory where the leading term in a derivative expansion of the bulk action is \cite{NS}
\bea \label{SSUSY}
 S_\ssB &=& - \int \exd^6x \sqrt{-g} \; \left[ \frac{1}{2\kappa_6^2} \; g^{\ssM\ssN} \left( \cR_{\ssM\ssN} + \partial_\ssM \phi \, \partial_\ssN \phi \right)  + \frac{2 \mfg^2}{\kappa_6^4} \; e^\phi  \right. \nn\\
 && \qquad\qquad\qquad\qquad\qquad \left.  + \frac{1}{12} \, e^{-2\phi} H_{\ssM\ssN\ssP} H^{\ssM \ssN \ssP} + \frac14 \, e^{-\phi} F_{\ssM\ssN} F^{\ssM \ssN} \right] \,,
\eea
where $H$ is a 3-form Kalb-Ramond field strength and $F$ is a 2-form Maxwell field strength and $\phi$ is a 6D scalar dilaton. $\kappa_6$ is the 6D gravitational coupling and $\mfg$ is the 6D gauge coupling for the Maxwell field $F_{\ssM\ssN}$. Notice this action scales as $S_\ssB \to \zeta^2 S_\ssB$ when the fields are scaled by
\be \label{6DScalingSym}
   g_{\ssM \ssN} \to \zeta \, g_{\ssM\ssN} \quad \hbox{and} \quad e^{-\phi} \to \zeta \, e^{-\phi} \quad \hbox{with} \quad H_{\ssM\ssN\ssP} \; \hbox{and} \;  F_{\ssM\ssN} \; \hbox{held fixed}
\ee
where $\zeta$ is a constant (and the same is also true for the fermionic part of the lagrangian). Although this is {\it not} a symmetry of the action it {\it is} a symmetry of the classical equations of motion. 

The leading terms in the derivative expansion of the action for each 4D brane is
\bea \label{SSUSYb}
 S_b = - \int_{W_b} \exd^4 x \sqrt{-\gamma} \; \left( \cT_b +\tfrac{1}{4!} \cA_b \, \varepsilon^{\mu\nu\lambda\rho} \cF_{\mu\nu\lambda\rho} + \cdots \right) \,,
\eea
where $W_b$ denotes the brane's world-volume and $\varepsilon_{\mu\nu\lambda\rho}$ is the volume form built from its induced metric $\gamma_{\mu\nu}$. Here the 4-form is the 6D dual of the Maxwell field:\footnote{The $\phi$-dependence in $\cF$ ensures that its Bianchi identity $\exd \cF = 0$ properly captures the Maxwell field equation implied by the lagrangian \pref{SSUSY}.} 
\be \label{6DMaxwellDual}
    \cF_{\ssM\ssN\ssP\ssQ} = \tfrac12 \epsilon_{\ssM\ssN\ssP\ssQ\ssR\ssS} \, e^{-\phi} F^{\ssR\ssS} \,,
\ee
where $\epsilon_{\ssM\ssN\ssP\ssQ\ssR\ssS}$ is the volume form built from the 6D metric. Notice that the lagrangian \pref{SSUSYb} scales in the same way as did $S_\ssB$ if $\cT_b$ and $\cA_b$ are $\phi$-independent (and so do not scale). 

The parameter $\cT_b$ can be interpreted physically as the brane's tension. The quantity $\cA_b$ measures the amount of Maxwell flux that is localized on the brane, in the sense that the $\cA_b$ term changes the Maxwell equation to become
\be
   \partial_m \Bigl[ e^{-\phi} \Bigl( \sqrt{g_2} \, F^{mn} - \sum_b \cA_b \, \epsilon^{mn} \, \delta^2(x-x_b) \Bigr) \Bigr] = 0 \,,
\ee
where $\epsilon_{mn}$ is the volume form for the extra-dimensional 2D geometry \cite{Companions}. This introduces localized magnetic flux at the position of each brane, and changes the flux-quantization condition to
\be \label{FluxQ2}
     \int_{M_2} F_{mn} +  \sum_b \cA_b \, \frac{\epsilon_{mn}}{\sqrt{g_2}}   = \frac{n}{\mfg} \,,
\ee
where $n$ is an integer. Notice that this condition does not break scale invariance (because $F_{\ssM\ssN}$ doesn't scale while $\epsilon_{mn}$ and $\sqrt{g_2}$ scale in the same way) {\it provided} that $\cA_b$ is independent of $\phi$. 

The simplest solution to the 6D field equations having maximal 4D symmetry arises when the two branes are identical, in which case the extra dimensions form a rugby ball \cite{SS, SLED} for which $H_{\ssM\ssN\ssP} = 0$, $\phi = \phi_0$ and
\be \label{2DmetricSS}
 \exd s^2 = \hat g_{\mu\nu} \, \exd x^\mu \, \exd x^\nu + \ell^2 \Bigl( \exd \theta^2 + \beta^2 \sin^2 \theta \, \exd \xi^2 \Bigr) e^{-\phi_0} \quad \hbox{and} \quad
 F_{\theta\xi} = Q \beta \ell^2 \sin \theta \,,
\ee
where $\hat g_{\mu\nu}$ is a maximally symmetric 4D geometry with curvature scalar $\widehat R$ and $\phi_0$, $Q$, $\beta$ and $\ell$ are constants. With this ansatz the field equations boil down to
\be \label{SUSYsoln}
 \frac{1}{\ell^2} = \kappa_6^2 Q^2 = \left( \frac{2 \mfg}{\kappa_6} \right)^2 \,, \quad   1-\beta = \frac{\kappa_6^2 \cT}{2\pi} \quad \hbox{and} \quad \widehat R = 0
 \,.
\ee

Notice that the value of $\phi_0$ is not determined by any of the field equations; a consequence of a classical scale invariance \pref{6DScalingSym}.  Furthermore the flux-quantization condition \pref{FluxQ2} is satisfied for this solution provided the integer is $n = \pm 1$ and the flux localized on the branes is
\be \label{FluxQ2a}
    \Phi_{\rm tot} := \sum_b  \cA_b   = \pm \frac{1 - \beta}{e_6} \,.
\ee
If $\cA_b$ were to depend on $\phi$ then flux quantization would break the scale invariance and \pref{FluxQ2a} can be used to determine the value of $\phi_0$ given a functional form for $\cT_b(\phi)$ and/or $\cA_b(\phi)$, as was explored in some detail in \cite{Burgess:2011mt}.

One more extra-dimensional property is required before discussing the theory's low-energy EFT, and the role played in it by the 4-form field $F_{\mu\nu\lambda\rho}$. The limit where $\cA_b$ and $\cT_b$ vanish describes a solution of the bulk field equations \cite{SS} for which the geometry \pref{2DmetricSS} admits a Killing spinor and so leaves a single 4D supersymmetry unbroken. The boundary condition at the brane breaks this supersymmetry for generic nonzero values for $\cT$ and $\cA$, but for a specific value of $\cA/\cT$ the brane's magnetic flux and tension are related by a BPS condition that allows the bulk supersymmetry to survive even for nonzero\footnote{For a precursor to this argument see \cite{precursor}.} $\cT$ \cite{SUSYRestored}. It does so because then the boundary conditions satisfied by the metric and Maxwell fields allow the extra-dimensional gauge and metric connections to be equal, and so restore the existence of a Killing spinor.

\subsubsection{The effective 4D point of view}

We next describe the effective 4D description of the above system appropriate below the KK scale. The field content of this EFT is the 4D metric, $g_{\mu\nu}$, corresponding to the 4D metric $\hat g_{\mu\nu}$ of the 6D theory; the 4D  scalar $\varphi$ corresponding to the modulus $\phi_0$ in the 6D theory and a 4-form $\cF_{\mu\nu\lambda\rho}$, corresponding in the 6D theory to the dual $F_{\mu\nu\lambda\rho}$ obtained from \pref{6DMaxwellDual} applied to the background Maxwell field $F_{mn}$ in the extra dimensions. 

Of these, the field $\cF_{\mu\nu\lambda\rho}$ might be a surprise because, unlike for the other fields, $F_{mn}$ does not represent a light degree of freedom in the compactification. But the background value of $F_{mn}$ is quantized through the relation \pref{FluxQ2} with low-energy implications because it explicitly appears in the low-energy lagrangian. It is the non-propagating field $\cF_{\mu\nu\lambda\rho}$ that brings the news about extra-dimensional flux quantization to the 4D EFT. A field is required (as opposed to simply having quantized couplings in the effective theory, for example) because quantization need not occur for all choices of vacua that the EFT can describe. The EFT must be as flexible in its response as is its microscopic UV completion.

Explicitly, the flux-quantization condition in 6D dualizes to give
\be
  F_{\mu\nu\lambda\rho} =  \left( \frac{ \cN + \zeta }{\cZ} \right) \, \epsilon_{\mu\nu\lambda\rho} \,,
\ee
where $\cN := 2\pi n/\mfg$ and $\zeta$ and $\cZ$ are explicitly specified in \cite{Burgess:2015lda} in terms of $\cT_b$ and $\cA_b$ and the extra-dimensional geometry (the details of which do not concern us here). 

The most general 4D effective lagrangian for the above fields at the two-derivative level is (in 4D Einstein frame)
\bea \label{EF4D}
  \cL_4 &=& - \sqrt{- g} \; \left\{ \frac{1}{2\kappa_4^2} \, g^{\mu\nu} \Bigl[ R_{\mu\nu} + Z_\varphi(\varphi) \, \partial_\mu \varphi \, \partial_\nu \varphi \Bigr] + V_4(\varphi) \right. \nn\\
  && \qquad\qquad\qquad \left. + \frac{1}{2\cdot 4!} \, Z_\ssF(\varphi) \, F_{\mu\nu\lambda\rho} F^{\mu\nu\lambda\rho} - \frac{1}{4!} \, \xi(\varphi) \, \epsilon^{\,\mu\nu\lambda\rho} F_{\mu\nu\lambda\rho} \right\} + \cL_{st4} \,,
\eea
where the surface term, $\cL_{st4}$, is required when manipulating the 4-form field (see Appendix \ref{App:Scalar-4form}). 

The functions $V_4(\varphi)$, $Z_\ssF(\varphi)$, $Z_\varphi(\varphi)$ and $\xi(\varphi)$ are chosen to match the predictions of the 4D EFT to the equivalent results computed using the full 6D theory. For instance requiring the equations of motion for $F_{\mu\nu\lambda\rho}$ and $\cF_{\mu\nu\lambda\rho}$ to agree in particular implies $\xi = \zeta$ and $Z_\ssF = \cZ$. Details of this matching are given in \cite{Burgess:2015lda} but for the present purposes what matters is what is found for the effective scalar potential, which determines both the vacuum value for $\varphi$ (if this exists) and the curvature of the 4D spacetime. 

What is important is the contribution to the 4D curvature proportional to $(\cN +\xi)^2$ as computed from back-reaction in the full 6D theory is precisely captured by a term in the potential found by integrating out $\cF_{\mu\nu\lambda\rho}$ in the 4D EFT. In particular it is the presence of this term that shows why the low-energy 4D potential vanishes once $\sum_b \kappa^2_6 \cT_b$ is related to $\sum_b \mfg \cA_b$, in precisely the way predicted by the BPS condition preserving supersymmetry in the 6D theory. 

\subsubsection{Coupling strengths}
\label{sssec:CouplingStrengths}

This higher-dimensional model provides a representative example for which the size of the extra contact term in the axion-matter couplings can be computed. This involves understanding the size of the scales $f$, $\cZ$ and $m$ in the lagrangian \pref{L2axionmassive}, repeated here (including metric couplings) for convenience of reference
\be \label{L2axionmassive2}
  \frac{ \cL_2(\axion)}{\sqrt{-g}}
  = -\frac{f^2}2  \Bigl( \partial_\mu \axion + J _\mu \Bigr) \Bigl( \partial^\mu \axion + J ^\mu \Bigr)  - \frac{1}{2\cZ^2}(m \axion - \cJ)^2  + \frac{1}{4!}\,  \axion \, \epsilon^{\mu\nu\lambda\rho}  \Omega_{\mu\nu\lambda\rho}  + \cL_{\rm st} \,.
\ee

To estimate the size of the $J^\mu$ coupling for the theory described in \S\ref{sssec:SalamSezgin} we use the fact that $B_{\mu\nu}$ arises as the massless KK mode involving the 4D components\footnote{This is often called the `univeral' axion because this particular zero mode relies for its existence on there being a nontrivial solution to $\nabla^2 u = 0$ within the extra dimensions for some scalar $u$, and $u = \hbox{constant}$ does the job for any compact space.} of $B_{\ssM\ssN}$. Because $B_{\mu\nu}$ and $C_{\mu\nu\lambda}$ are both `bulk' fields their kinetic energies both involve the volume of the extra dimensions and so a canonically normalized field like $\widehat G_{\ssM\ssN\ssP}$ in 6D is related to a canonically normalized field $G_{\mu\nu\lambda}$ in 4D by $\widehat G_{\mu\nu\lambda} = r^{-1} G_{\mu\nu\lambda}$, where $r$ is the extra-dimensional radius. Since, for canonically normalized 4D fields \S\ref{sec:DualAxions} shows $f  \simeq \cZ^{-1} \simeq \cO(1)$ the axion mass \pref{axionmass} becomes $m_a = m/(f\cZ) \simeq m$ and $J^\mu$ has dimension (mass)${}^2$ in fundamental units. 

The rescaling of the fields when passing from 6D to 4D implies a similar rescaling of the 6D current since in higher-dimensional theories the matter couplings arise as couplings of higher-dimensional fields to a current or scalar involving Standard Model fields on a 4D brane. For instance, for a canonically normalized 3-form field in 6D a coupling $g\, \widehat G \wedge \hat J$ (with $g$ a dimensionless coupling) involves a current $\hat J^\mu$ with dimension (mass). Typical SM currents like $E^\mu = \ol\psi \Gamma^\mu \psi$, where $\psi$ is 4D spin-half field and $\Gamma^\mu$ is some matrix of dimensionless coefficients, have higher dimension than this and so the natural choice uses $\hat J^\mu = E^\mu/M_6^2$ where the higher-dimensional Planck scale $M_6$ lies at multi-TeV energies in \S\ref{SSUSY}. This leads on dimensional reduction to $g\, G \wedge J$ where $J^\mu = g E^\mu/(M_6^2r) = g E^\mu /\Mp$, after $\widehat G$ is rescaled to remain canonical in 4D (and using the relation $M_p = M_6^2 r$ between 4D and 6D Planck masses). Similarly, for a coupling $\hat \cJ_d H_{\mu\nu\lambda\rho}$ with $\hat \cJ_d$ a SM singlet scalar operator with mass dimension $d$ that breaks the axion shift symmetry, dimensional reduction gives $\cJ_d = \hat \cJ_d M_6^{3-d}/M_p$, 

For a QCD-type axion $E^\mu$ is the axial current for a PQ type symmetry \cite{PQ} and $\Omega$ in \pref{L2axionmassive2} is the 4-form built from gluon field strengths while a dimension-4 term $\hat\cJ_4$ arises from the quark Yukawa couplings. In this case it is usually convenient to perform a symmetry rotation to set the $\axion\, \Omega$ term to zero, at the expense of changing the form of $\cJ$. But more broadly $E^\mu$ could instead be the baryon number current, or some other gauge-invariant quantity. 

For macroscopic applications the current $E^\mu$ can usually be written $E^\mu = n U^\mu$ where $n$ is the local average density of the appropriate charge and $U^\mu$ is the 4-velocity of the macroscopic fluid's local rest frame. For flat geometries the components of the 4-velocity in a general frame are $U^0 = \gamma$ and $\bfU =  \gamma \, \bfv$ where $\gamma = (1 - \bfv^2)^{-1/2}$ and so $E^0 = \gamma n$ and $\bfE = \gamma n \bfv$.

\subsection{Relevance to Dark Energy?}

This section describes two ways in which non-propagating 4-forms in 4D arise within discussions of the cosmological constant problem. The first of these gives a brief summary of unimodular gravity and why (in our opinion) this has so far not made convincing progress on the problem. We argue that the 4-forms that descend from extra-dimensional models -- like the one discussed in \S\ref{sssec:SalamSezgin} -- can help with the failings of unimodular gravity because their connection to extra-dimensional supersymmetry can help provide the important missing ingredient. 

Although nothing in this paper requires one to believe any particular approach to the cosmological constant problem, this section closes with a brief discussion of relaxation models -- such as the Yoga approach presented in \cite{Burgess:2021obw} (which captures the low-energy 4D limit of the extra-dimensional systems). We do so because it illustrates another potential difference between generic scalar axions and axions that arise as duals: they differ in the extent with which their scalar potential is suppressed by any physics that makes progress with the cosmological constant problem. 

In particular, both Yoga models and their extra-dimensional predecessors (like \cite{SLED}) suggest that axions that arise in the traditional way -- {\it i.e.}~as phases of scalar order parameters in the particle-physics sector -- tend to have their scalar potential suppressed by the same physics that is responsible for suppressing the Dark Energy density itself. When this happens it changes the relationship between the axion's cosmological properties (like masses) and the underlying symmetry-breaking scales, making the axion much lighter than would have naively been expected. The same tends {\it not} to happen for axions whose potentials arise as the duals of 4-form kinetic terms. 

\subsubsection{4-forms and unimodular gravity}

Unimodular gravity \cite{Unimodular1, UnimodularAction, Unruh:1988in, Finkelstein:2000pg, Alvarez:2005iy, Shaposhnikov:2008xb, Smolin:2009ti, Padilla:2014yea} (see \cite{Alvarez:2023utn} for a recent review) was perhaps the earliest variant of general relativity aimed specifically at understanding the nature of the cosmological constant. It is defined as the theory obtained by restricting the variation of the metric in GR to those that have fixed volume: $\sqrt{-g} = w$ (with units usually chosen so that $w = 1$). Such a condition is not diffeomorphism invariant (if $w$ is fixed), being invariant only under volume-preserving diffeomorphisms.  

Unimodular gravity is also closely related to theories of 4-form potentials. Although this connection often seems academic from the perspective that 4-form form fields do not describe propagating particles, it might acquire a new light considering the differences between naturalness criteria formulated on either side of a duality relationship discussed in \S\ref{ssec:NaturalnessDual}.  

Varying the Einstein-Hilbert (plus matter) action only over metrics $g_{\mu\nu}$ whose determinant is not varied leads to the trace-free Einstein equations
\be \label{unimodeq}
    G_{\mu\nu} - \tfrac14 G \, g_{\mu\nu} + \kappa^2 \Bigl( T_{\mu\nu} - \tfrac14 \, T \, g_{\mu\nu} \Bigr) = 0 \,,
\ee
where $G_{\mu\nu} = R_{\mu\nu} - \tfrac12  R \,g_{\mu\nu}$ is the metric's Einstein tensor and $T_{\mu\nu}$ is the stress-energy tensor for any matter fields. Quantities without indices denote the corresponding covariant traces: $G := g^{\mu\nu} G_{\mu\nu}$ and $T := g^{\mu\nu} T_{\mu\nu}$. 

At face value eq.~\pref{unimodeq} leaves the curvature scalar $R := g^{\mu\nu} R_{\mu\nu}$ unconstrained, but taking its divergence and using the Bianchi identity $\nabla^\mu G_{\mu\nu} = 0$ implies $\partial_\mu G + \kappa^2 \partial_\mu T = 0$ and so 
\be
    \partial_\mu R = \kappa^2 \partial_\mu T \,,
\ee
provided we assume the matter stress-energy is covariantly conserved: $\nabla^\mu T_{\mu\nu} = 0$. This integrates to give $R = \kappa^2 T -2c$ where $c$ is an integration constant. Combined with the trace-free equation \pref{unimodeq} this implies the complete Einstein tensor therefore satisfies
\be
   G_{\mu\nu} + c \, g_{\mu\nu} + \kappa^2 T_{\mu\nu} = 0 \,,
\ee
which are the usual Einstein equations with the integration constant $c$ playing the role of the cosmological constant. 

Unimodular gravity has been argued to be an appealing approach to the cosmological constant problem because it moves the problem from a term in the action to something that is instead determined by initial (or boundary) conditions. Any other contributions to the scalar potential remain present, however, and can be functions of time ({\it e.g.}~if the universe were to undergo a phase transition) so it is not clear why the integration constant should cancel them and (if it does) why it should do so now.

A connection to 4-forms is natural if the constraint on $\sqrt{-g}$ is written in form language. In 4D the metric defines the volume 4-form $\mfv = - \frac{1}{4!} \epsilon_{\mu\nu\lambda\rho} \, \exd x^\mu \wedge \exd x^\nu \wedge \exd x^\lambda \wedge \exd x^\rho$ where $\epsilon_{\mu\nu\lambda\rho}$ is the metric's Levi-Civita tensor (which is completely antisymmetric, and in our conventions $\epsilon_{0123} = - \sqrt{-g}$ so $v_{0123} = +\sqrt{-g}$). Any other 4-form $\mfw = \frac{1}{4!} w_{\mu\nu\lambda\rho} \, \exd x^\mu \wedge \exd x^\nu \wedge \exd x^\lambda \wedge \exd x^\rho$ has components $\pm w$ where $w$ transforms the same way as does $\sqrt{-g}$. Implementing a constraint on the determinant of the metric can therefore be accomplished by constraining the volume form to be equal to a particular fixed 4-form. (See {\it e.g.}~\cite{Guendelman:1999tb} for an alternate use of multiple measures in the cosmological constant problem.)

This can be done concretely within a path integral by adding a lagrange-multiplier term to the action of the form
\be
  S \to S + \int s \Bigl( \mfw - \mfv \Bigr) \,,
\ee
where $s(x)$ is a scalar field and the integral is over all of 4D spacetime. Functional integration over $s(x)$ gives a functional delta function that everywhere imposes the constraint $\mfw = \mfv$. In components this corresponds to 
\be \label{linearw}
  \frac{\cL}{\sqrt{-g}} \to \frac{\cL}{\sqrt{-g}} - \frac{s}{4!} \,  \epsilon^{\mu\nu\lambda\rho} \Bigl(  w_{\mu\nu\lambda\rho} + \epsilon_{\mu\nu\lambda\rho} \Bigr) =  \frac{\cL}{\sqrt{-g}} + s  \left( 1 - \frac{w}{\sqrt{-g}} \right) \,,
\ee
where $w_{0123} = w$ defines $w$.

If the metric integral is performed before the $s$ integral and evaluated semiclassically then because the metric is now unconstrained the saddle point occurs when the standard Einstein equations are satisfied
\be
  G_{\mu\nu} - s \, g_{\mu\nu} + \kappa^2 T_{\mu\nu} = 0 \,.
\ee
Because $\sqrt{-g} \, \epsilon^{\mu\nu\lambda\rho}$ is metric independent the term involving $w$ does not contribute at all to the Einstein equations, but the term involving $\mfv$ contributes to them as an $s$-dependent contribution to the stress energy of the form $T_{\mu\nu}^{(s)} \propto s\, g_{\mu\nu}$. The field equation for $s$ is not obtained by varying it in the action (this instead just imposes the constraint $\sqrt{-g} = w$). It is instead found by taking the divergence of the Einstein equation, since this implies $\partial_\mu s = 0$ (once we use the Bianchi identity and conservation of the rest of $T_{\mu\nu}$). Again we see a cosmological constant emerge as an integration constant when solving to get $s = c$.

\subsubsection{More generic 4-forms}

Terms like those appearing in \pref{linearw} naturally arise when 4-form fields appear in the path integral, though with a big difference: one integrates over the form fields rather than treating them as being specified quantities. The action is also then usually more general than simply the linear dependence on $\mfw$ that was assumed above. A typical lagrangian that might be encountered at the two-derivative level would have the form
\be \label{Dualstarter}
 \frac{ \cL}{\sqrt{-g}} = \frac{\cL_0}{\sqrt{-g}}  - \Bigl[ V + \tfrac{1}{2\cdot 4!} \cZ^2 H^{\mu\nu\lambda\rho} H_{\mu\nu\lambda\rho} + \tfrac{1}{4!} \xi \, \epsilon^{\mu\nu\lambda\rho} H_{\mu\nu\lambda\rho} \Bigr]  
\ee
where $\cL_0$ contains the Einstein-Hilbert terms and the contributions of other matter fields to the action, though the scalar potential $V$ for these other fields is separated out and made explicit. The potential $V$ and the other coefficients $\cZ$ and $\xi$ are scalar functions that can depend on any other scalar fields.  

If we integrate out the 4-form field by performing a saddle point evaluation of the integral over its 3-form potential $C_{\nu\lambda\rho}$ the saddle point is
\be \label{xCfieldeq}
    \partial_\mu\Bigl[ \sqrt{-g} \Bigl( \cZ^2 H^{\mu\nu\lambda\rho} + \xi \epsilon^{\mu\nu\lambda\rho} \Bigr) \Bigr]= 0 \,,
\ee
This has as solution
\be \label{xAppHSaddle2}
      \cZ^2 \cH^{\mu\nu\lambda\rho} + (\xi - k) \epsilon^{\mu\nu\lambda\rho}   = 0 \,,
\ee
where $k$ is an arbitrary constant. Using this in the action and being careful with surface terms  -- see Appendix \ref{App:Scalar-4form} for more details -- leads to
\be \label{DualVshift}
 \frac{ \cL}{\sqrt{-g}} = \frac{\cL_0}{\sqrt{-g}}  -   \widetilde V  \qquad \hbox{where} \qquad 
   \widetilde V = V + \frac{(\xi - k)^2}{2\cZ^2} \,.
\ee

The net effect of the 4-form dynamics is simply to add a strictly non-negative contribution to the potential, in the form of a perfect square. The size of the square added also depends on an integration constant $k$. This positivity makes 4-form contributions similar to auxiliary field contributions to scalar potentials in supersymmetric theories, and as summarized in \cite{Bielleman:2015ina} this is often not a coincidence.

If the coefficient $\xi$ depends on other scalar fields then the fact that the 4-form terms are a square means they are always minimized at zero, if this is possible for some scalar field configuration. Unimodular gravity emerges as a special case in this framework as the limit $\cZ \ll 1$ together with the special case $\xi = s$ and a term included in $V$ that is linear in $s$: $V \to V - s$, for some scalar field $s$. In this case the limit $\cZ \to 0$ makes the potential steep enough to overwhelm any kinetic energies for $s$, and so the $(\xi - k)^2/\cZ^2$ term in $\widetilde V$ has the effect of trapping $s$ at the value $s = k$. 

\subsubsection{Relaxation models}

As mentioned above, although 4D 4-form fields can be related to unimodular gravity this does not in itself make them a useful solution to the cosmological constant problem. In particular, why should any integration constant remove a previous quantum contribution to $V$ that was already present and does not involve the 4-form field? We remark in passing that extra-dimensional models like the one discussed in \S\ref{sssec:SalamSezgin} make some progress on this by showing how under some circumstances the 4-form contribution to the potential (involving $\cA_b$) can cancel the form-independent terms (involving $\cT_b$) -- see \cite{Burgess:2015lda} for details. 

From the higher-dimensional point of view this cancellation occurs because for a specific relation between $\cA_b$ and $\cT_b$ the near-brane boundary conditions allow the bulk spin and gauge connections to agree and this in turn allows a Killing spinor to exist in the bulk, implying that one 4D supersymmetry becomes restored in the bulk even in the brane's presence. From the 4D point of view the bulk looks like a supersymmetric dark sector with gravitational couplings \cite{Burgess:2004yq, Burgess:2021juk} and so the cancellation occurs when supersymmetry becomes restored in the dark sector despite the presence of non-supersymmetric brane-localized particles ({\it i.e.}~the particle physics sector). In extra-dimensional models (like \cite{SLED}) this cancellation is dynamical and adaptive as the extra-dimensional geometry responds to changes to the values of the localized stress-energy.  

The adaptiveness of extra-dimensional models makes them good illustration of relaxation mechanisms for suppressing the cosmological constant (though of course not all relaxation mechanisms are extra-dimensional \cite{Graham:2015cka}). In extra dimensional models it is the entire brane tension that ultimately gets suppressed regardless of its origin and field dependence. This includes any axion potential if the axion is a regular 4D scalar field that arises as the phase of an order parameter that is localized on the brane, leading to a suppression of the axion mass relative to naive expectations. 

For instance, in the Yoga approach presented in \cite{Burgess:2021obw} (which captures the low-energy 4D limit of the extra-dimensional systems), the small size of the Dark Energy density arises because the low-energy theory has a dilaton field $\sigma$, in terms of which Standard Model particle masses are order $M_\EW \sim M_p \, e^{-\sigma/2}$ and the scalar potential has the form $V(\sigma) =M_p^4 U \, e^{-4\sigma}$ (where $U$ is a polynomial function of $\sigma$). The coefficients in $U$ can easily ensure a minimum for $\sigma \sim 60$ large enough to ensure that particle masses and the minimum value of $V$ are both sufficiently small. 

Now comes the main point: in these models an ordinary brane-based axion's potential also appears in $U$ (rather than additively) and so is also suppressed when the value of $\sigma$ is large. When this happens it changes the relationship between the axion's cosmological properties (like masses) and the underlying symmetry-breaking scales, making the axion much lighter than would have naively been expected. For the present purposes what is important is the same tends {\it not} to happen for axions whose potentials arise as the duals of 4-form kinetic terms, because these arise in the extra-dimensional bulk rather than localized on the brane.  

\section{Discussion}
\label{sec:Discussion}

In this article we revisit some quantum aspects of $p$-form dualities in QFT in order to emphasize the differences between generic systems and those that admit duals.
In particular we argue that antisymmetric tensor field theories in 4D carry practical information that can make their dual theory differ in interesting ways from garden-variety scalars. 

We consider two ways in which the existence of dual formulations can be interesting: the dual can differ on the size of corrections to semiclassical calculations, or one side of a dual pair can bring sensitivity to topology. There are also two reasons why corrections can differ in size between a theory and its dual, one of which is well-known (the duality could swap strong with weak couplings) and the other less so (the duality maps derivative to non-derivative interactions, which contribute differently in low-energy expansions). We mostly focus here on illustrating this latter case, using as a vehicle the duality between form fields and a massive scalar.  We explore several of its implications, with the following highlights:
\begin{itemize}
\item We review the standard powercounting arguments in the two dual theories and confirm that they go through in much the same way on both sides. In particular control over the semiclassical approximation on both sides comes from the low-energy expansion itself, similar to what happens for gravity.  
\item For massive scalars we repeat the arguments that show that duality maps the zero-derivative scalar interactions coming from the scalar potential map onto derivative couplings on the form-field side. This poses a potential problem for power-counting because it mixes up terms at different orders of the derivative expansion.
\item In order to agree with the dual theory at low energies, the couplings in the scalar potential are suppressed in the precise way required to restore the consistency of the low-energy limit.  In particular the scalar field $\axion$ always appears in the scalar potential multiplied by its mass, so $V(\axion) = V(m \axion)$. If higher-derivatives on the form-field side are suppressed by powers of a UV mass $M$ then the coefficient of a term like $\mfa^n$ in the potential is suppresed by powers of $(m/M)^n$ instead of being of order one, effectively ensuring that scalar couplings count as derivatives so far as the low-energy expansion is concerned.
\item Standard topological arguments imply that the axion is periodic, so in the limit $M\rightarrow \infty$ the potential for the axion is the superposition of quadratics centered about the discrete minima (as opposed to a trigonometric function). Trigonometric functions can be obtained by duality but correspond on the dual side to cases where the field strengths are comparable to $M$ and so higher-derivatives are not negligible. When this is so semiclassical expansions on the form-field side of the duality are not justified in the usual way by the derivative expansion. 
\item When the kinetic coefficient $f^2$ of an axion field has physical content (such as if it is a function of other fields) then duality maps $f\rightarrow 1/f$, sometimes allowing smaller values of $f$ than standard to make sense. This may have interesting physical implications.
\item If the axion couples to matter through just a $\partial_\mu \axion J^\mu$ interaction then the dual coupling is a combination of a term $\exd B \wedge J$ and a $J^\mu J_\mu$ term. Conversely, if the form-field coupling is $\exd B \wedge J$ then the coupling on the scalar side is $(\partial_\mu \mfa +J_\mu)^2$. For situations where the UV theory `picks' the form-field side (such as within extra-dimensional models) and so interactions arise with independent couplings then the unexpected contact $J^\mu J_\mu$ term should be expected on the scalar side. This might also have interesting physical implications.                                                                  
\end{itemize}

There are also other situations where derivative and non-derivative couplings get interchanged under duality. One of these is the standard relation between $f(R)$ theories and scalar-tensor theories, for which curvatures on the $f(R)$ side are mapped onto non-derivative terms in the scalar potential on the scalar field side. Unlike the Kalb-Ramond case ordinary low-energy power-counting arguments do not apply on the $f(R)$ side, particularly in cases where the linear term in $R$ competes with higher powers of curvature (such as in Starobinsky inflation). In this case a clean statement about the control over semiclassical methods must await the development of new tools beyond the usual derivative expansion. Until then the scalar field side of the duality is the only one in which corrections are under control (which can effectively be used to define what is meant by the $f(R)$ theory).

Our analysis also clarifies several misunderstandings in the literature. In particular, in \cite{Hell:2021wzm} it was claimed that previous (correct) claims in the literature about the duality between massive $p$ forms and massive $D-p-1$ forms in $D$ dimensions must be incorrect since perturbation of both systems by a particular quartic coupling gave different results. But we have shown how duality maps non-derivative couplings to derivative couplings, so there is no reason to expect equivalence of the duals if they are {\it both} deformed using non-derivative interactions. Once a theory is deformed on one side the duality relation itself tells you what the equivalent deformation is on the other side. Ref.~\cite{Hell:2021wzm} is correct that the assumed perturbations give different results but this is only because the two assumed deformations are not dual to one another.  

We also remark that there is more than one way to give a mass to a 2-form Kalb-Ramond potential $B_{\mu\nu}$ in four dimensions. In $D$ spacetime dimensions a massless $p$-form gauge potential (plus its ghosts) propagates a total of $s=\frac12(D-p)(D-p-1)$ degrees of freedom, which in 4D becomes $s=1$ for a 2-form (Kalb-Ramond) field and $s=0$ for a 3-form field. The corresponding result for a massive $p$-form potential in $D$ dimensions is $s = \frac12 (D-p+1)(D-p)$, which in 4D becomes $s=3$ for a massive 2-form potential and $s = 1$ for a massive 3-form. 

The option where the single spin state of a massless 2-form combines with the two helicities of a massless spin-one particle in 4D is the form-field version of the ordinary Higgs mechanism -- where a spin-1 particle `eats' a spinless one. This is always possible because the shift symmetry of the scalar obtained by dualizing a Kalb-Ramond field ensures it can be regarded as a Goldstone boson. This effect has been established explicitly in \cite{Burgess:1995kp} and discussed at length in \cite{Quevedo:1996uu}, including the general $p$-form case and the standard Stuckelberg mechanism in which a gauge field eats the axion field and becomes massive. But there is nothing that says that breaking the 2-form gauge symmetry in 4D must give 3 massive spin states because the relevant mechanism might instead involve a 2-form and 3-form potential combining to give a single massive spin state (along the lines we mostly study in this paper). 
  
The `dual is different' in our title refers to the different structure that dual scalars can have relative to their generic counterparts, not that the theories involved cease to be dual. Our particular interest is in how this difference can change one's priors about what is natural or unnatural in the scalar potential (priors that are often invoked when selecting models for cosmology). Duality has not stopped surprising us over several decades and there will surely be more surprises to come.

\section*{Acknowledgements}
We thank Brian Batell, Carsten van de Bruck, Pedro Ferreira, Rich Holman, Anton Hook, Antonio Iovino, Ted Jacobson, Simon Lin,  Daniel Mata-Pacheco, Shoy Ouseph, Pellegrino Piantadosi, Mario Ramos-Hamud, Sergey Sibiryakov, Costas Skordis, Adam Smith, Gonzalo Villa and Luis Zapata for useful discussions and Stefan Theisen for drawing our attention to reference \cite{Hell:2021wzm}. CB's research was partially supported by funds from the Natural Sciences and Engineering Research Council (NSERC) of Canada. Research at the Perimeter Institute is supported in part by the Government of Canada through NSERC and by the Province of Ontario through MRI. The research of FQ has been funded by a grant from New York University Abu Dhabi.

\begin{appendix}

\section{4-forms and boundary terms}
\label{App:Scalar-4form}

This section is meant as a reminder of how scalar/4-form interactions work and in particular how there are surface terms that can make integrating out the 4-form subtle (for details see \S3.1 and \S3.2 of \cite{Burgess:2015lda}).

Consider the lagrangian encountered in the duality argument
\be \label{AppDualstarter}
  \cL = - \sqrt{-g} \Bigl[ \tfrac12 f^2 (\partial_\mu \axion + J_\mu)(\partial^\mu \axion + J^\mu)  + V + \tfrac{1}{2\cdot 4!} \cZ^2 H^{\mu\nu\lambda\rho} H_{\mu\nu\lambda\rho} + \tfrac{1}{4!} \xi \, \epsilon^{\mu\nu\lambda\rho} H_{\mu\nu\lambda\rho} \Bigr]  
\ee
where we (to start with) do not worry about a possible surface term $\cL_{\rm st}$, and the coefficients $\cZ$, $\xi$ can be scalar functions of other fields.

\subsubsection*{Integrating $H$}

Consider the following steps. First integrate out $H_{\mu\nu\lambda\rho}$ as an unconstrained integral. This has saddle point $H_{\mu\nu\lambda\rho} = \cH_{\mu\nu\lambda\rho}$ where
\be \label{AppHSaddle}
      \cZ^2 \cH^{\mu\nu\lambda\rho} + \xi \epsilon^{\mu\nu\lambda\rho}   = 0 \,,
\ee
so integrating out gives
   \bea
  \cL
  &=& - \sqrt{-g} \left[ \tfrac12 f^2 (\partial_\mu \axion + J_\mu)(\partial^\mu \axion + J^\mu) +V - \tfrac{1}{2\cdot 4!  } \frac{\xi^2}{ \cZ^2}  \epsilon^{\mu\nu\lambda\rho} \epsilon_{\mu\nu\lambda\rho}  \right]  \\
  &=& - \sqrt{-g} \left[ \tfrac12 f^2 (\partial_\mu \axion + J_\mu)(\partial^\mu \axion + J^\mu) +V +  \frac{\xi^2}{2 \cZ^2}    \right] \,. \nn
\eea
This shows that the 4-form sector modifies the potential by adding to it
\be
   V \to V + \frac{\xi^2}{2\cZ^2} \,.
\ee
\subsubsection*{Integrating $C$}

If we instead integrate out the 4-form field by performing a saddle point evaluation of the integral over $C_{\nu\lambda\rho}$ the saddle point this time satisfies
\be \label{Cfieldeq}
    \partial_\mu\Bigl[ \sqrt{-g} \Bigl( \cZ^2 H^{\mu\nu\lambda\rho} + \xi \epsilon^{\mu\nu\lambda\rho} \Bigr) \Bigr]= 0 \,,
\ee
This is weaker than \pref{AppHSaddle} because $k \epsilon_{\mu\nu\lambda\rho}$ has vanishing covariant divergence where $k$ is an arbitrary constant. So \pref{AppHSaddle} gets replaced by
\be \label{AppHSaddle2}
      \cZ^2 \cH^{\mu\nu\lambda\rho} + (\xi - k) \epsilon^{\mu\nu\lambda\rho}   = 0 \,.
\ee
This changes the evaluation of the action because
\bea \label{notquiteright}
      - \sqrt{-g} \Bigl[  \tfrac{1}{2\cdot 4!} \cZ^2 \cH^{\mu\nu\lambda\rho} \cH_{\mu\nu\lambda\rho} + \tfrac{1}{4!} \xi \, \epsilon^{\mu\nu\lambda\rho} \cH_{\mu\nu\lambda\rho} \Bigr]  &=&   - \frac{1}{4! \cZ^2}  \sqrt{-g} \Bigl[   \tfrac{1}{2} (\xi - k)^2 - \xi (\xi - k) \Bigr] \epsilon^{\mu\nu\lambda\rho} \epsilon_{\mu\nu\lambda\rho}  \nn\\
      &=&     \frac{1}{ \cZ^2}  \sqrt{-g} \Bigl[   \tfrac{1}{2} (\xi - k)^2 - \xi (\xi - k) \Bigr]  \\
      &=& -  \frac{1}{ \cZ^2}  \sqrt{-g} \Bigl[   \tfrac{1}{2} (\xi - k)^2 + k (\xi - k) \Bigr]  \,.\nn
\eea
The first term is precisely what is needed to have the potential become
\be \label{Vshift}
   V \to V + \frac{(\xi - k)^2}{2\cZ^2} \,,
\ee
but the last term apparently ruins this when $k \neq 0$. Eq.~\pref{Vshift} is nonetheless the right answer however, as can be seen below once the saddle point for $H$ is used in the axion and metric field equations. 

The way to see this directly in the action as well requires being more careful with the surface terms. Consider, for instance, the surface term
\be
   \cL_{\rm st4} =  \tfrac{1}{3!} \partial_\mu \Bigl[ \sqrt{-g} \Bigl(    \cZ^2 H^{\mu\nu\lambda\rho}  +   \xi \, \epsilon^{\mu\nu\lambda\rho}\Bigr)   C_{\nu\lambda\rho} \Bigr]  \,,
\ee
which can be seen to be gauge invariant (despite the explicit $C_{\nu\lambda\rho}$) after integration by parts. Such terms introduce an important ambiguity in determining how the action depends on fields after integrating out $C$ because these surface terms contribute in an important way to the action evaluated at solutions to the bulk equations of motion. Once evaluated on shell $\cL_{\rm st4}$ becomes
\be 
   \cL_{\rm st4} =  \tfrac{1}{4!} \Bigl[ \sqrt{-g} \Bigl(    \cZ^2 \cH^{\mu\nu\lambda\rho}  +   \xi \, \epsilon^{\mu\nu\lambda\rho}\Bigr)  \cH_{\mu\nu\lambda\rho} \Bigr] =   - \tfrac{1}{4!}  \frac{k(\xi-k)}{\cZ^2} \Bigl[ \sqrt{-g}  \; \epsilon^{\mu\nu\lambda\rho}   \epsilon_{\mu\nu\lambda\rho} \Bigr]  =       \frac{k(\xi-k)}{ \cZ^2}  \sqrt{-g}     \,,
\ee 
where the first equality uses \pref{Cfieldeq} and the second equality uses \pref{AppHSaddle2}. This is precisely what is needed to cancel the extra term in \pref{notquiteright}, leaving the $H$ contribution to the on-shell action as
\be
      - \sqrt{-g} \Bigl[  \tfrac{1}{2\cdot 4!} \cZ^2 \cH^{\mu\nu\lambda\rho} \cH_{\mu\nu\lambda\rho} + \tfrac{1}{4!} \xi \, \epsilon^{\mu\nu\lambda\rho} \cH_{\mu\nu\lambda\rho} \Bigr]  + \cL_{\rm st4} =   - \sqrt{-g} \Bigl[ - \tfrac{1}{2\cdot 4!} \cZ^2 \cH^{\mu\nu\lambda\rho} \cH_{\mu\nu\lambda\rho}   \Bigr] 
\ee
as required to reproduce \pref{Vshift}. 

\subsubsection*{Arguments within the equations of motion}

To check the shift in $V$ independent of surface terms, check the field equations. Suppose that $\xi$ is a function of $\axion$, so the axion equation that follows from \pref{AppDualstarter} is
\be
    \frac{1}{\sqrt{-g}} \partial_\mu \Bigl[ \sqrt{-g} \; f^2 (\partial^\mu \axion + J^\mu) \Bigr] -   V'  - \tfrac{1}{4!} \xi' \, \epsilon^{\mu\nu\lambda\rho} H_{\mu\nu\lambda\rho}  = 0 \,,
\ee
so evaluating this at \pref{AppHSaddle2} then gives
\bea
   0 &=& \frac{1}{\sqrt{-g}} \partial_\mu \Bigl[ \sqrt{-g} \; f^2 (\partial^\mu \axion + J^\mu)  \Bigr] -   V'  - \tfrac{1}{4!} \xi' \, \epsilon^{\mu\nu\lambda\rho} \cH_{\mu\nu\lambda\rho}  \nn\\
   &=& \frac{1}{\sqrt{-g}} \partial_\mu \Bigl[ \sqrt{-g} \; f^2 (\partial^\mu \axion + J^\mu)  \Bigr] -   V'  + \tfrac{1}{4!} \frac{(\xi - k)}{\cZ^2}  \xi' \, \epsilon^{\mu\nu\lambda\rho} \epsilon_{\mu\nu\lambda\rho}   \\
   &=& \frac{1}{\sqrt{-g}} \partial_\mu \Bigl[ \sqrt{-g} \; f^2 (\partial^\mu \axion + J^\mu)  \Bigr] -   V'  - \frac{(\xi - k)}{\cZ^2}  \xi' \, \,,\nn
\eea
precisely as expected if the potential were shifted as in \pref{Vshift}.

On the other hand, for the Einstein equations what counts is the stress energy. Differentiating the action with respect to the metric gives
\bea
   - \frac{1}{\sqrt{-g}} \frac{\delta S}{\delta g_{\mu\nu}} &=& \frac12 \Bigl[ \tfrac12 f^2 (\partial_\alpha \axion + J_\alpha)(\partial^\alpha \axion + J^\alpha)  + V + \tfrac{1}{2\cdot 4!} \cZ^2 H^{\alpha\beta\lambda\rho} H_{\alpha\beta\lambda\rho}   \Bigr] g^{\mu\nu}  \nn\\
   && \qquad \qquad \qquad -  \Bigl[ \tfrac12 f^2 (\partial^\mu \axion + J^\mu)(\partial^\nu \axion + J^\nu)   + \tfrac{4}{2\cdot 4!} \cZ^2 H^{\mu\beta\lambda\rho} {H^\nu}_{\beta\lambda\rho}  \Bigr] 
\eea
and so 
the stress energy becomes
\bea
  T^{\mu\nu} =  \frac{2}{\sqrt{-g}} \frac{\delta S}{\delta g_{\mu\nu}} &=&   f^2 (\partial^\mu \axion + J^\mu)(\partial^\nu \axion + J^\nu)   + \tfrac{1}{3!} \cZ^2 H^{\mu\beta\lambda\rho} {H^\nu}_{\beta\lambda\rho}   \nn\\
   && \qquad   - \Bigl[ \tfrac12 f^2 (\partial_\alpha \axion + J_\alpha)(\partial^\alpha \axion + J^\alpha)  + V + \tfrac{1}{2\cdot 4!} \cZ^2 H^{\alpha\beta\lambda\rho} H_{\alpha\beta\lambda\rho}   \Bigr] g^{\mu\nu} \,.
\eea
Evaluating this at the saddle point \pref{AppHSaddle2} using
\bea
   \tfrac{1}{3!} \cZ^2 \cH^{\mu\beta\lambda\rho} {\cH^\nu}_{\beta\lambda\rho}  -  \tfrac{1}{2\cdot 4!} \cZ^2 \cH^{\alpha\beta\lambda\rho} \cH_{\alpha\beta\lambda\rho}   g^{\mu\nu} &=&  \frac{(\xi - k)^2}{\cZ^2} \Bigl[   \tfrac{1}{3!}  \epsilon^{\mu\beta\lambda\rho} {\epsilon^\nu}_{\beta\lambda\rho}  -  \tfrac{1}{2\cdot 4!}  \epsilon^{\alpha\beta\lambda\rho} \epsilon_{\alpha\beta\lambda\rho}   g^{\mu\nu}  \Bigr]  \nn\\
   &=& \frac{(\xi - k)^2}{\cZ^2} \Bigl[ -  \  g^{\mu\nu} +  \tfrac{1}{2}    g^{\mu\nu}  \Bigr] 
    = - \frac{(\xi - k)^2}{2\cZ^2}   g^{\mu\nu}  
\eea
then gives
\be \label{AppScalarTmn}
  T^{\mu\nu} =   f^2 (\partial^\mu \axion + J^\mu)(\partial^\nu \axion + J^\nu)     - \left[ \tfrac12 f^2 (\partial_\alpha \axion + J_\alpha)(\partial^\alpha \axion + J^\alpha)  + V  + \frac{(\xi - k)^2}{2\cZ^2}  \right] g^{\mu\nu} \,.
\ee
This again argues that the 4-form field contributes to the potential by an amount given by \pref{Vshift}.

\section{Driven oscillations}
\label{App:DrivenOsc}

This appendix records how axion oscillations are driven by the presence of matter when the current appearing in the interaction is {\it not} conserved.

\subsubsection{Oscillating axions (small scales)}

Consider the solution corresponding to a homogeneously oscillating axion in the presence of background matter, $J^0 = \lambda n$, such as would be expected to apply over astrophysical scales if the axion is dark matter. Since this is a classical solution it is possible to track what it looks like on the dual side to describe what the semiclassical background solution is in the dual variables. We use a quadratic potential since this suffices to capture the dual at the two-derivative level (and to describe small oscillations). 

On the scalar side the axion field equation \pref{axioneq} in flat space (weak gravitational fields) specializes for homogeneous and isotropic current distributions $J^0 = J^0(t) = \lambda n(t)$, $J^i = 0$ and homogeneous solutions $\axion(t)$ to
\be \label{axioneq2}
\ddot \axion + m^2 \axion = \lambda \dot n   \,,
\ee
where we drop $\Omega$ and set $f = \cZ = 1$ and assume $\cJ$ is constant (and then shift $\axion$ to set it to zero). The current term has the effect of driving the usual axion oscillations whenever the background density is time-dependent. The general solution to \pref{axioneq2} is
\bea
   \axion_c(t) 
   &=&  \axion_0 \, \cos(m t) + \frac{v_0}{m} \, \sin(m t) + \frac{\lambda}{m}\int_{0}^t \exd \tau \, \dot n(\tau) \, \sin[m (t - \tau)] \nn \\
   &=&   \axion_0 \, \cos(m t) + \left( \frac{v_0 - \lambda n_0}{m} \right) \sin(m t) + \lambda \int_{0}^t \exd \tau \, n(\tau) \cos[m (t - \tau)]\,, \nn
\eea
where $\axion_0 = \axion(0)$ and $v_0 = \dot \axion(0)$ are integration constants and $n_0 := n(0)$. Although this returns the standard monochromatic axion oscillation in regions of constant density in regions where $n(t)$ changes with time the driving implied by the matter interaction introduces new frequencies. 

The current-axion interaction changes the energy of this oscillation, because the equation of motion \pref{axioneq2} implies
\be
 \rho = \tfrac12 (\dot \axion - \lambda n)^2 + \tfrac12 m^2 \axion^2
\ee
satisfies
\be
 \dot \rho = \lambda m^2 n \, \axion_c \,,
\ee
and so integrated over $N$ oscillations this gives 
\bea \label{axionenergygrowth}
   \rho(T_\ssN = 2\pi N/m) &=& \rho(0) +  \lambda^2 m \int_0^{T_\ssN} \exd t \, n(t) \, \int_0^t \exd \tau \, \dot n(\tau) \, \sin[m (t - \tau)] \nn\\
   &=&  \rho(0) +  \lambda^2 m^2 \int_0^{T_\ssN} \exd t \, n(t) \, \int_0^t \exd \tau \, n(\tau) \cos[m (t - \tau)] \,. 
\eea

For instance, if the matter density is increasing with a rate proportional to the density then we would expect $\dot n = \Gamma n$. For astrophysical processes the time scale for this is much longer than the axion oscillation time so $\Gamma \ll m$. If we follow the energy transfer for $N \gg 1$ oscillations but for a total time much smaller than $1/\Gamma$ then $\Gamma T_\ssN \ll 1$ and we can approximate $n(t) \simeq n_0$ and $\dot n \simeq \Gamma n_0$. With these assumptions \pref{axionenergygrowth} gives a secular growth in axion energy of size 
\be
   \rho_\ssN - \rho_0 \simeq  \lambda^2n_0^2  \, \Gamma T_\ssN = \frac{2\pi N \lambda^2 n_0^2\, \Gamma}{m} = \frac{g^2 N G_\ssN   n_0^2\, \Gamma}{4m}\,,
\ee 
where the last equality uses the estimate $\lambda = g/\Mp$ given in \S\ref{sssec:CouplingStrengths} above and $G_\ssN = (8\pi \Mp^2)^{-1}$ is Newton's gravitational constant. For times longer than this the back-reaction to the gravitational field and onto the matter evolution cannot be neglected. 


\subsubsection*{Solution in dual variables}

We pause here briefly to record the form this solution takes on the form-field side. Using the above solution in \pref{HSaddle} implies
\be \label{HSaddle2}
   \cH_{\mu\nu\lambda\rho} =  m\, \axion_c(t) \, \epsilon_{\mu\nu\lambda\rho}
\ee
while \pref{GSaddlem} implies
\be \label{GSaddlem2}
    \widehat \cG_{\mu\nu\lambda} = - \epsilon_{\mu\nu\lambda\rho}  (\partial^\rho \axion_c + J^\rho)  =  \epsilon_{\mu\nu\lambda 0}  (\dot \axion_c + J_0)
    \,,
\ee
and so $\widehat \cG_{0ij} = 0$ and $\widehat \cG_{ijk} = (\dot \axion_c + J_0) \, \epsilon_{ijk}$ (since our convention is $\epsilon^{0ijk} = +\epsilon^{ijk}$). As a check $(\exd \widehat \cG)_{0ijk} = \partial_t(\dot \axion_c + J_0) \epsilon_{ijk}$ and so \pref{axioneq2} implies $\exd \widehat \cG = m \cH$, as required. In unitary gauge this amounts to giving $C_{ijk}$ because $\widehat \cG_{ijk} = m C_{ijk}$ and so we see
\be
 C_{0ij} = 0 \quad \hbox{and} \quad C_{ijk} =\frac{1}{m} ( \dot \axion_c + J_0) \, \epsilon_{ijk} \qquad \hbox{(unitary gauge)} \,.
\ee

\end{appendix}

\end{document}